\documentclass[11pt]{article}
\usepackage{graphicx} 
\usepackage{setspace}
\usepackage{lscape}
\usepackage{quoting} 
\usepackage{eurosym}
\usepackage{mathtools}
\usepackage{amsmath}
\usepackage{amssymb}
\usepackage{textgreek}
\usepackage{rotating}
\usepackage{multirow}
\usepackage{color}
\usepackage{xcolor}
\usepackage{cancel}
\usepackage{comment}
\usepackage{dcolumn}
\usepackage{hyperref}
\usepackage{wrapfig}
\usepackage{pdflscape}
\usepackage{booktabs,caption}
\usepackage[flushleft]{threeparttable}
\makeatletter
\let\xx@thm\@thm
\AtBeginDocument{\let\@thm\xx@thm}
\makeatother

\hypersetup{
    colorlinks = true,
    linkbordercolor = {white},
    linkcolor = {blue}}

\quotingsetup{font=small}

\title{{\LARGE \bf \textcolor{blue}{From the historical Roman road network \\[-2ex] to modern infrastructure in Italy}
}\\
\vspace{0.4cm}
}

\author{Luca De Benedictis\footnote{  DED - University of Macerata; Rossi-Doria Centre, University Roma Tre; and Luiss. Email: luca.debenedictis@unimc.it.}\  \  \   Vania Licio\footnote{ Corresponding author. DEPS - University of Siena and CRENoS. Email: vania.licio@unisi.it.} \  \  \ Anna Maria Pinna\footnote{  DSEA - University of Cagliari and CRENoS. Email: ampinna@unica.it.}  }

\begin{document}

\maketitle \thispagestyle{empty}

\begin{abstract}
\setlength{\baselineskip}{10.5pt} 

\vspace{0.2in}

\noindent \noindent {An integrated and widespread road system, like the one built during the Roman Empire in Italy, plays an important role today in facilitating the construction of new infrastructure. This paper investigates the historical path of Roman roads as main determinant of both motorways and railways in the country. The empirical analysis shows how the modern Italian transport infrastructure followed the path traced in ancient times by the Romans in constructing their roads. Being paved and connecting Italy from North to South, consular trajectories lasted in time, representing the starting physical capital for developing the new transport networks.}

\vspace{2.0cm} \noindent

\end{abstract}

\newpage
\setcounter{page}{1}

\newpage

\begin{quoting} ``{\itshape And what was said by Homer, `The Earth was common to all', you (Rome) have made a reality, by surveying the whole inhabited world, by bridging rivers, by cutting carriage roads through the mountains, by filling deserts with stationes, and by civilising everything with your way of life and good order}" Aelius Aristides Orat.26.101
\end{quoting}

\section{Introduction}
\label{sec_Intro}

Transport infrastructure is of major importance. Its direct and indirect impact on growth and economic development has been widely discussed in the literature.  However, there are still some gray areas, like the effect played by historical transport projects on modern physical capital. Within the Italian territory, the main mean of transport for moving goods in space is today through the use of roads.\footnote{ In 2016, 18 percent of the value of extra-EU28 exports in goods was by road. In Italy, it was 21 percent (Source: \url{http://ec.europa.eu/eurostat/statistics-explained/index.php?title=International_trade_in_goods_by_mode_of_transport}). Within Europe, road freight transport is predominant and represents approximately three-quarters of the total. In Italy, this share is even greater.} Lagging territories, like the  southern Italian regions - distant from the centers of economic activity - corroborate the need to invest in transport infrastructure as the engine of stimulating growth.\footnote{ See \hyperlink{roberts2020}{Roberts \textit{et al.} (2020)} for a quantitative review of the literature aimed at estimating the impacts of large transport infrastructure projects: transport investments may be beneficial for welfare, equity and social inclusion, but some heterogeneity can exists at the subnational level.} 
This study explores the role of the Roman roads in Italy, investigating whether the current transport infrastructure benefited from the existence of a paved network in the past.

Italy represents an ideal case: it is where the historical Roman road network, one of the largest investments in infrastructure in history, was densest and where its expansion began. The Roman Empire had its core in Italy and roads were built throughout the peninsula. Rome can be thought as the `point source outbreak'  of the Roman conquest pattern, which took several centuries to unfold.  This aspect introduces an element that has to do with the economic fortunes of Italy today. The Italian economic and social dualism can be attributed to historical episodes and dominations that occurred following the collapse of the Roman Empire (\hyperlink{carla2017}{Carl\`a-Uhink, 2017}). Since the Middle Ages, the peninsula has been ruled by several foreign dominators, who were quite heterogeneous within the current unified territory, both in cultural and administrative terms. Today, the central government has the main influence in determining institutions. However, national regulations work differently in the North and the South of the country, suggesting that specific local factors affect the institution's functioning.\footnote{ \hyperlink{crescenzi2016}{Crescenzi \textit{et al.} (2016)} stress the importance of the quality of regional government for the positive economic returns of transport investments at the local level.} The idea at the heart of this work is that Roman roads have positively affected current transport systems, regardless of the variety of historical paths within the Italian territory.
 
This paper follows the strand of research that quantifies the long-term effects of historical events on current development (\hyperlink{nunn2009}{Nunn, 2009}),  and the line of investigation of \hyperlink{temin2013}{Temin (2013)}, \hyperlink{michaels2018}{Michaels and Rauch (2018)}, \hyperlink{wahl2017}{Wahl (2017)}, \hyperlink{dalgaard2018}{Dalgaard \textit{et al.} (2018)} and \hyperlink{flueckiger2022}{Flueckiger \textit{et al.} (2022)}. The evidence here provided embraces two dimensions: the persistent effect of history and the mechanism linking past with present. The unit of analysis is the 10x10 grid cell level with a comprehensive list of controls to capture all those geographic, urban, and historical factors crucial in determining both old and modern transport networks. 

This paper is also related to the literature on the rise, development and growth of Italian cities, which started with \hyperlink{malanima2005}{Malanima (2005)}, who describes the long-term urbanization process in Italy, referring to labor productivity forces and balances between rural and town areas. \hyperlink{bosker2013}{Bosker \textit{et al.} (2013)} allow for the nature of a city's geographical location, where the fact of being connected to a major Roman road or a Roman hub makes the difference when accounting for the possibility that smaller cities become larger population centers. \hyperlink{percoco2013}{Percoco (2013)} widens the view by using historical city characteristics to instrument firm and employment density and estimating their role on income growth.

Historians have reported how Romans built their roads, particularly the major (consular) ones, to conduct their military campaigns. During the Roman Empire goods were transported mainly by sea, while the mobility of people, for purposes other than military, was virtually non existent. This paper provides novel empirical evidence: the presence of pre-Roman cities and amenities along with geography have been related to the length of Roman roads. Romans did not use major roads for connecting existing cities. Instead, they were motivated by finding solutions for overcoming geographical barriers, the first point of the paper. The second result refers to how the modern infrastructure has benefited from the existence of ancient roads, instead. Nowadays, there are more railways and motorways in areas where the Roman network was denser. 

The paper is organized as follows. Section \ref{section_2} places the analysis in the literature. Section \ref{section_3} illustrates the available data on Roman roads, and discusses the exogeneity of their path and the least cost trajectories. Section \ref{section_4} offers the empirical validation of historians' argument that Romans built their roads not intending to connect economic centers. Section \ref{section_5} investigates the link between ancient and modern infrastructure. Section \ref{section_6} concludes.

\section{\label{section_2}     The persistent effect of historical infrastructures}

In the last twenty years a new strand of literature, focused on the influence of history on various aspects of the economy today,  has emerged. As summarized by \hyperlink{nunn2009}{Nunn (2009)} and  \hyperlink{michalopoulos2017}{Michalopoulos and Papaioannou (2017)}, the `new economic history' literature -- started by collecting data on specific historical episodes (e.g. colonialism in \hyperlink{laporta1997}{La Porta \textit{\textit{et al.}}, 1997} and \hyperlink{acemoglu2001}{Acemoglu \textit{\textit{et al.}}, 2001}) and providing evidence of their long-lasting effect on modern economic development -- rapidly evolved in several directions. The historical epochs under scrutiny - from the Neolithic to Nazi Germany -- and the geographical expansion were one of the first directions investigated. In a second stage, the research ``[...] \textit{moved from asking whether history matters to asking why history matters}'' (\hyperlink{nunn2009}{Nunn, 2009, p.66}) and several studies focused on the mechanism linking the past to contemporary outcomes, exploring the channels of causality in identification-based empirical analyses.

The focus on historical infrastructures accounts for a large share of this literature. Recent research has shown an interest in the effect of great historical transportation infrastructure projects and their expansion on reducing trade costs, on enhancing productivity and on increasing the level of real income in the trading regions involved,  as in the contributions by \hyperlink{donaldson2018}{Donaldson (2018)}, \hyperlink{donaldson2016}{Donaldson and Hornbeck (2016)}, \hyperlink{jedwab2017}{Jedwab \textit{et al.} (2017)}, \hyperlink{berger2017}{Berger and Enflo (2017)}. The first two papers address railway expansion in colonial India and the U.S. respectively, the first also providing a measure of the share of total welfare gains. The other two papers provide similar evidence for Kenya and Sweden.\footnote{ Other works address other means of transport.  \hyperlink{fajgelbaum2014}{Fajgelbaum and Redding (2014)} look at Argentina and the reduction of international transport costs generated in the late nineteenth century by the introduction of large steamships. \hyperlink{volpe2014}{Volpe Martincus \textit{et al.} (2014)} analyze the case of Peru and use the Inca road network (built by the Inca Empire before 1530) as an instrument for the current road infrastructure.} 

Differently from other episodes in history, the civilization that started from Rome in 753 B.C. stood out for its prolonged and extended traits. The lasting marks, left in terms of performing institutions, urbanization patterns, and the development of a market economy, led several studies to focus on the positive effect attributable to the Roman domination (\hyperlink{bosker2017}{Bosker and Buringh, 2017}; \hyperlink{buringh2012}{Buringh \textit{et al.}, 2012}; \hyperlink{bosker2013}{Bosker \textit{et al.}, 2013}; \hyperlink{michaels2018}{Michaels and Rauch, 2018}); recently, the Roman road infrastructure attracted the attention of researchers. In the paper by \hyperlink{wahl2017}{Wahl (2017)}, the presence of ancient Roman roads is instrumental in dividing the area that corresponds to contemporary Germany into a Roman and a non-Roman part. The \emph{Limes Germanicus wall} is used as a geographical discontinuity to test whether the formerly Roman part of Germany shows greater nighttime luminosity than the non-Roman one. The transmission mechanism is attributed to the enduring effect of the Roman road network,  fostering city growth and denser infrastructure. \hyperlink{dalgaard2018}{Dalgaard \textit{et al.} (2018)}, instead, use the network of roads constructed during the Roman Empire to demonstrate the provision of public goods as a channel of persistence of economic development. The result is corroborated by comparing the European region with the Middle-East and North Africa territories. Since in Africa the wheel was substituted by camels, Roman roads were not maintained and cannot explain current economic performance. In Europe, instead, Roman road maintenance offers a valid proof of the persistence of infrastructure over time.  \hyperlink{flueckiger2022}{Flueckiger \textit{et al.} (2022)} document a lasting economic integration in Western Europe due to the ancient Roman transport infrastructure and its effect on the modern connectivity.


\section{\label{section_3} The Roman road network}

The Roman road network started to spread simultaneously with the expansion of Romans in the IV century B.C. The main reason for constructing paved suburban roads was purely military: the need to rapidly deploy troops to the insecure borders of the Empire.\footnote{ Consular roads were not built either for trade purposes, largely managed by navigation across the Mediterranean Sea, or for civilian transportation  (\hyperlink{chevallier1976}{Chevallier, 1976}).} Of course, Romans built roads also for connecting settlements and cities, but this paper demonstrates how this is not the case for major (consular) roads.

The earliest strategic consular Roman road was the \textit{Via Appia} (the Appian Way), named after the Roman censor Appius Claudius Caecus and constructed in 312 B.C. in a south-easterly direction from Rome to Capua (close to Naples), to guarantee fresh troops for the war against the Samnites. It was constructed as an assembly of straight segments as the easiest and most convenient way to move troops and supplies. Subsequently, the Appian Way, the \textit{regina viarum}, was extended to reach Beneventum, and Tarentum and finally Brundisium, on the Adriatic coast. 

After the \textit{Via Appia}, miles and miles of roads were built.\footnote{ Starting from the city of Rome, the expansion of the road network covered six centuries and three continents (Europe, Africa, Asia).} 
Figure \ref{figureA1} - included in \nameref{AppendixA} - provides a representation of the main consular Roman roads that crossed the Italian territory. In Italy, the network of Roman roads covered the entire peninsula, including the two main islands (Sicily and Sardinia). There is no other country or region where the Roman infrastructure was dense and widespread as in Italy. Romans used to divide their roads into primary (major) and secondary (minor). Major Roman roads led directly to Rome - if placed in Italy - or to Italy - if running outside current Italian borders. They were linked to each other, composing a network of military or main roads. Secondary roads, instead, linked military roads to cities and settlements.

\subsection{\label{subsection_3.1} Data}

The raw data on the Roman road network was digitized by \hyperlink{mccormick2013}{McCormick \textit{et al.} (2013)} on the basis of the \hyperlink{talbert2000}{Barrington Atlas of the Greek and Roman World (2000)}.\footnote{ The data, in shape file format, allows spatial analysis for the Roman and medieval worlds using a Geographic Information System (GIS) coding.}
The Digital Atlas of Roman and Medieval Civilizations (DARMC) includes 7,154 segments of ancient Roman roads existing at the peak of the Empire, corresponding with the death of Trajan (117 A.D.). Each segment is uniquely identified and roads are composed of many segments.\footnote{ As an example, the \textit{Via Appia} is composed of 67 different segments. Roads are not classified as such and have to be reconstructed assembling the different segments. For brevity, from now on, the terms `road' and `road segment' will be used interchangeably.
} 
The network covers 36 countries over Europe, Africa, and Asia, and road segments are classified according to their class of importance (e.g. major and minor) and certainty (e.g. certain and uncertain).\footnote{ Certainty refers to the path followed by the road segment. All segments are always certain in their existence and in their Roman origin; what makes a road `uncertain' is the imprecision in the georeferentiation of the path followed by the road: some stretches of roads got destroyed or abandoned through the ages and for some of them there is neither archaeological nor historical evidence.} Figures \ref{figureA2} and \ref{figureA2} - included in \nameref{AppendixA} - provide a representation of Roman roads according to importance and certainty.

The road network covers a total distance of 192,861 kilometers. The length of the Roman road infrastructure at the Italian 10 x 10 \textit{km} grid cell level - the observation unit of this paper - is computed by \hyperlink{licio2021}{Licio (2021)}. Figure \ref{figure1} shows the old road system for Italy, which comprises 10 percent of the entire network: 1,817 segments for a total of 19,593 kilometers in 2,068 out of 5,111 10 x 10 \textit{km} grid cells. 1,359 have major roads, 1,032 minor roads: 315 both minor and major, 717 only minor and 1,044 only major (consular) Roman roads.

\begin{center}
[Figure \ref{figure1}]
\end{center}

\subsection{\label{subsection_3.2} Exogeneity}

As generally applies, the exogeneity of the Roman road infrastructure must be placed under scrutiny.\footnote{ A considerable amount of time elapses between planning a road and its actual completion (\hyperlink{brooks2009}{Brooks and Hummels, 2009}). From this perspective, road infrastructure can be considered as an exogenous variable.
The case of old infrastructure would appear to be different. \hyperlink{donaldson2018}{Donaldson (2018)} argues that the effect of historical transportation infrastructure is characterized by a potential simultaneity problem: roads and railways are often constructed to connect regions already active in trade, while inter-regional trade relations are often forged after the construction of infrastructure or road improvements.}
Roman roads might have been constructed for military purposes (major roads) but it cannot be ruled out that some of them and some minor roads were built to reach economically prosperous and flourishing territories, and that these conditions could well have lasted up to the present day.
\footnote{ \hyperlink{chevallier1976}{Chevallier (1976, p. 116)} points out that ``{\itshape As a rule, earlier sites were avoided by Roman roads, especially the great Imperial highways, which were unconcerned with local interests and small settlements. [...] The road often attracted the village, but when the ancient road itineraries name a civitas, it does not mean that the route went through the town itself: occasionally it simply skirts its territory}". \hyperlink{bosker2013}{Bosker \textit{et al.} (2013)} support the view that the reverse causality is not an issue in the case of Roman roads, since they favored the subsequent expansion of urban centers in those territories where roads passed through, rather than being constructed for already existing settlements.} 

\subsubsection{\label{subsubsection_3.2.1}  Why did Romans build roads? The `military reason'}

As reported in the Dictionary of Greek and Roman Antiquities, ``{\itshape The public road-system of the Romans was thoroughly military in its aims and spirit: it was designed to unite and consolidate the conquests of the Roman people, whether within or without the limits of Italy proper}" (\hyperlink{smith1890}{Smith, 1890}). Even after construction, it had no significant immediate economic impact, since the cheaper modes of goods' transport in that historical epoch were by river or sea (\hyperlink{finley1973}{Finley, 1973}). More specifically, as \hyperlink{laurence1999}{Laurence (1999)} clearly explains, roads were planned and designed to provide troops with the essential means in terms of subsistence and support, to guarantee an efficient repositioning and to facilitate armies' movements. Because of this original purpose, roads were straight, as level as possible, often stone-paved, cambered for drainage, equipped with safe stops along the way.\footnote{ The `military reason' is also strongly supported by the Latin literature. ``{\itshape After having pacified Liguria, Aemelius had his army build a road from Piacenza to Rimini to join the via Flaminia}" (Livy, 59 B.C.-17 A.D.). In his `Encyclopedia of antiquities, and elements of archaeology, classical and medieval,' \hyperlink{fosbroke1843}{Fosbroke (1843)} reports that the Anglo-Saxon ancestors named the Roman roads `military ways' and that they thought the construction of small roads had more military utility than large ones. Chevallier emphasizes the importance of the army's role in the case of main roads. He remarks that ``{\itshape [...] the majority of main roads were pioneered by military operations. For example, on its return from the first Samnite war (343-40), the Roman army did not come back along the via Latina, but followed the coast through the territory of Aurunci, thus blazing the trail of the Appia on a line that had already been known to traders, at least since the hegemony of Etruria. In the early third century, operations against the Umbrians of Mevania and Narnia and against the Senones took into account the route that became the Flaminia. Great strategic roads were built by the military in Gaul under Agrippa from BC 16-13 in Dalmatia and Pannonia under Tiberius from AD 6-9, in the Rhineland and the Danube valley under Claudius, and in Asia Minor under Flavians}" (\hyperlink{chevallier1976}{Chevallier, 1976, p. 85}).}

The direction of the \textit{Via Appia}\footnote{ See \hyperlink{ber2003}{Berechman (2003)} for a recent and in-depth description of the economics of the \textit{Via Appia}.} is a concrete example of its military purpose and how the ultimate aim to reach some strategic territories resulted in a long road that passed through areas of absolutely no interest to Romans, but that, nonetheless, benefited from the presence of the road.\footnote{ The Romans decided to build their first road south-easterly, although the economic development of the time was concentrated in the southwestern part of Italy in those territories corresponding today to the NUTS2 regions of Campania, Calabria and Sicily.}

\subsubsection{\label{subsubsection_3.2.2} The case of the \textit{Via Appia} and the Greek target}

The \textit{Via Appia} was the first large strategic consular road. It connected Rome to Brundisium (modern Brindisi, Apulia). Started in 312 B.C, the road had the original tactical purpose to allow troops to be deployed outside the region of Rome during the Samnite Wars.\footnote{ Chambers's Encyclopedia, Vol. 1, p. 490.}  It was constructed in segments, following the progress of military campaigns, and was completed in 191 B.C., when it reached Brindisi \hyperlink{ber2003}{(Berechman, 2003)}. 

\begin{center}
[Figure \ref{figure2}]
\end{center}

During that time, the Roman Empire was comprised, as shown in the upper part of Figure \ref{figure2}, in those territories belonging to the Latine League\footnote{ The Latine League is a term coined by modern historians, that identifies a coalition of villages and tribes settled in central Italy, surrounding Rome and that had the primary role in guaranteeing the mutual protection against external enemies (\hyperlink{cornell1995}{Cornell, 1995}).} and corresponding, today, to the provinces of Rome and Latina. Between 500 and 400 B.C. the Romans had already defeated their neighbors in central Italy (the Etruscans, Latins,  Sabines, Lavinii, Tusculi, Aequi, Volsci, Aurunci and the Veientes), where a small area was under their control. At the end of the fifth century B.C. the Italian peninsula was under the control of the Celts and the Gauls in the North, the Romans in the central-western part, the Samnites and the Greek colonies (Magna Graecia) in the South. It was precisely at that time, that the Romans decided to build the first section of the \textit{Via Appia} (lower part of Figure \ref{figure3}) and started to show an interest in the southern part of Italy. 
Also the Samnites, an Italic population living in southern-central Italy, were interested in those territories. At first, the Romans and the Samnites concluded a non-aggression pact, agreeing to expand their possessions in different directions, but this treaty was irremediably broken when these directions clashed. The Romans' intention was, first, to expand their territories in southern Italy (upper part of Figure \ref{figure3}), to obtain new lands for the growing Roman population and to enter into commercial relationships with the Greek merchants (\hyperlink{musti1990}{Musti, 1990}),  but, later, the ultimate challenge was the conquest of the Magna Graecia and extending their control over the Mediterranean Sea, where most of the trade occurred (Figure \ref{figure3}). 

\begin{center}
[Figure \ref{figure3}]
\end{center}

\begin{center}
[Figure \ref{figure4}]
\end{center}

In 238 B.C., the Romans controlled the entire central and southern parts of the Italian peninsula (upper part of Figure \ref{figure3}), including the three main Mediterranean islands (Sicily, Sardinia and Corsica); at the same time, the \textit{Via Appia} (lower part of Figure \ref{figure3}) was extended southeastwards, reaching Brindisi. The Romans aimed to expand northwards (in those territories under the control of the Celts), into Gaul, Spain, North Africa and into Greece (upper part of Figure \ref{figure4}). The step wise extension of the \textit{Via Appia} to Brindisi meant that the troops could sail from this port when they later conquered Epirus, landing on the Macedonian coast thanks to ally ports along the opposite coastline, like Durres. The lower part of  Figure \ref{figure4} shows how the \textit{Via Appia} facilitated the conquest of Greece.

Three facts emerge from the above description: 1) the instrumental role of roads in the military conquest of new territories; 2) the development and expansion of roads by strategic points: the Romans built new road segments starting from tactical cities or outposts (\emph{Stationes}); 3) the step wise construction of roads, with a view to future expansion. This suggests that some territories were crossed by Roman roads although the Romans themselves had no economic, military or tactical interest in those areas. In other terms, those territories benefited from the presence of Roman roads merely by chance, because they were situated midway between the origin of a road's segment and its strategic destination.

\subsubsection{\label{subsubsection_3.2.3} The case of the \textit{Via Annia} and the Magna Graecia}

Further and clear evidence that roads served as a means to exercise military control over the territory, and that they were designed in successive sections, is the \textit{Via Annia}.

The \textit{Via Annia}, also known as Via Popilia or Via Popilia Lenate or Via Capua-Regium,\footnote{ The name is still being debated since it is not certain which Roman consul was responsible for the construction of the road.} was the most important road in southern Italy. Built in 132 B.C. (200 years after the \textit{Via Appia}) at the request of the Roman judiciary and, with its 475 kilometers in length, it had the precise function of connecting Capua to Regium (modern Reggio Calabria) so as to guarantee military power in the \textit{Civitas foederata Regium}.\footnote{ In the second century B.C., Romans had already secured the entire South of Italy and they needed a road that easily connected Rome with the foot of the peninsula.} 

\begin{center}
[Figure \ref{figure5}]
\end{center}

The \textit{Via Annia} covered a tortuous path. As shown in Figure \ref{figure5}, the road ran along the western coastline of Campania and Calabria. Starting from Capua, it arrived at Salernum (modern Salerno) first and then at Regium, touching several Roman \textit{Stationes}: Nuceria (modern Nocera), Moranum (modern Morano), Cosentia (modern Cosenza), Valentia (modern Vibo Valentia). These stations were all located to the West. At that time, economic development and urbanization was concentrated along the eastern Calabrian coastline, where the former Greek colonies were located. Sibari, Crotone and Locri were large and important centers for trade. 
However, the route was not intended to connect all those municipalities.

Two main indications emerge from the construction of the \textit{Via Annia}: 1) the Roman concept behind road construction is closely related with creating a system of roads that connects strategic military points rather than a network among the major urban centers, and this is true of both major and minor roads; 2) roads were conceived as the assembly of linear segments, the easiest and fastest way to join two strategic points. 

\subsubsection{\label{subsubsection_3.2.4}  How did Romans build roads? The `engineering reason'}

One remarkable engineering feature of the Roman network was its straight roads: the Romans drew straight lines between two strategic locations and built the road as segments connected to one another. \hyperlink{cornell1982}{Cornell and Matthews (1982)} point out that the first step in road construction consisted of marking as straight as possible a path with stakes and furrows, employing sightlines as measuring tools.\footnote{ The Romans preferred direct and straight roads, because with that outline it was easier to avoid ambuscades and human settlements. Moreover, straight roads were easier to secure (\hyperlink{gleason2013}{Gleason, 2013}). As suggested by \hyperlink{poulter2010}{Poulter (2010)} and as remarked by \hyperlink{bishop2014}{Bishop (2014)}, the Romans often did guard the beginning and end of the road; garrisons were typically placed at the top of a hill, and the road came along as the segment of a paved route connecting two garrisons. \hyperlink{vonhagen1967}{Von Hagen (1967)}, on the constitution of a mobile civilization throughout the continent, argues how this has been possible thanks to well-engineered and straight roads.}

The rule that guided Romans in building roads is clearly explained by \hyperlink{lopez1956}{Lopez (1956, p. 17)} who describes ``{\itshape That the network of roads should be convenient and economic was none of their\footnote{ \hyperlink{lopez1956}{Lopez (1956)} refers to the Romans.} business. That is why the Romans built narrow, precipitous roads along the mountain crests rather than the valley bottoms, sometimes driving straight for their goal over gradients of one in five}". Also \hyperlink{margary1973}{Margary (1973)} remarks that, in order to achieve as straight a line as possible, Romans built roads with steep slopes or passing through mountainous terrains. \hyperlink{bishop2014}{Bishop (2014)}, referring to Britain, quoting \hyperlink{hindle1998}{Hindle (1998)} and \hyperlink{welfare1995}{Welfare and Swan (1995)}, emphasizes that long straight sections were a typical feature of the major Roman roads. However, even where variations in terrain morphology existed, the roads were still built in straight lines. Most of the non-major Roman roads exhibit some deviations from the main path. These variations in the course of the road were typically short and, rather than being curvy, they were subject to a change in the degree of the layout. This represents the typical feature that distinguishes Roman infrastructure from  modern infrastructure.\footnote{ In light of this, \hyperlink{bishop2014}{Bishop (2014)} refers to the Roman roads as `{\itshape surveyed roads}' which originate from a geometric-linear perspective in conceiving the network. Current roads are, instead, in the words of \hyperlink{bishop2014}{Bishop (2014)}, more linked to the `{\itshape line of desire,}' since there is no geometric outline behind the planning of the network, but rather a preference to follow the shape of nature.} 
Giving credit to historians' arguments, the straightness of the roads\footnote{ The lines connecting two points in space, such as two main cities, are the focal point of the identification strategy of a strand of economic literature that started with the work of \hyperlink{banerjee2012}{Banerjee \textit{et al.} (2012)}. Straight lines capture the way the first modern transportation infrastructure was constructed, which by definition cannot be influenced by the actual level of development, whereas the infrastructure developed afterwards was built along historical routes. According to this reasoning, straight lines can be used as the optimal tool for guaranteeing access to infrastructure and to disentangle the areas that benefited from the infrastructure, due to their proximity to the line (treated areas), from those that did not, because of the distance (non-treated areas).}

\subsection{\label{subsection_3.3} Linear, geography-based, real paths: an exercise}

If historical sources describe the straightness of Roman roads, linearity is also a desired feature from a mere geometrical point of view: straight lines are the shortest way to connect two points. Assuming flat terrains or not demanding geography, linear paths between targeted nodes minimize transport construction costs.
Straight lines have been used by the economic literature to exogenously identify the access to transportation. This literature has also relied on computing geographical-based least cost paths, based on elevation and land cover, to produce optimal networks in terms of cost minimization and not endogenous trajectories aimed at connecting cities  (\hyperlink{faber2014}{Faber, 2014}). In these terms, linear and geography driven trajectories represent the most efficient and exogenous routes in transportation projects. 

This section provides a simple exercise by building two hypothetical roads - a linear path and a geography-based one - and comparing them with the real Roman road path. Six main major Roman roads have been selected\footnote{\textit{Via Appia}, \textit{Via Aemilia}, \textit{Via Flaminia}, \textit{Via Aurelia}, \textit{Via Postumia}, \textit{Via Valeria}.} so that to represent different geographies (plain, hill, mountain, coastal), areas (north, center, south) and strategic territories (Magna Graecia, Gaul, Etruria, etc.).  

After selecting the consular roads, the second step consisted in drawing the corresponding linear path for each Roman road, i.e. the road resulting from straight lines connecting points and providing the direction Romans were committed to accomplish. To this aim, historical sources have been used to identify the main strategic/military points and targeted nodes that, step by step, ruled the construction of the road in ancient times. Nodes have been then connected using segments and creating a linear path.

The geography-based least cost path to be compared with the ancient Roman road has been obtained by using the raster of the topography of the territory for computing the cost surface in a buffer area of 10 \textit{km} around the linear path. Given the trajectory conceived by the Roman constructors, the most suitable geography-based path has been therefore calculated.\footnote{ Computations have been performed using QGIS. Results are confirmed using ArcGIS.}

Figures from \ref{figureA4} to \ref{figureA9} - included in \nameref{AppendixA} - provide for each selected Roman road the corresponding linear road and the geographical least cost path. The maps reveal how the shape of consular Roman roads conforms more to a line than to the most geographical advantageous layout. The finding is evident in all maps: see the section from Tarentum (modern Taranto) to Brundisium (modern Brindisi) on the Appian Way (Figure \ref{figureA4}); the complete path of the \textit{Via Aemilia} (Figure \ref{figureA5}); the segment between Vada Volaterrana and Luna (modern Luni) on the \textit{Via Aurelia}  (Figure \ref{figureA6}); several portions of the \textit{Via Flaminia}  (Figure \ref{figureA7}); the complete correspondence between the \textit{Via Postumia} and a straight line in the section between Bedriacum (modern Calvatone) and Verona (Figure \ref{figureA8}). For coastal roads like the \textit{Via Aurelia} and the \textit{Via Valeria} (Figure \ref{figureA9}), instead, the similarity between the ancient consular road and the least cost path is more pronounced since the plan of the road was constrained by a limited area. 

\section{\label{section_4} Roman roads, geography and pre-Roman cities}

Even if a direct economic reason driving the construction of the consular roads has been excluded by historians, geography and market access measures will be tested as possible factors influencing the construction of the ancient transport infrastructure. Roman roads may have been endogenously built where the morphology of the terrain permitted and/or according to economic reasoning of reaching further territories. 

Landform can shape both the within-country spatial distribution of road infrastructure and economic activity (\hyperlink{ramcharan2009}{Ramcharan, 2009}) and, if so, it represents a potential unobserved factor correlated with both road building and economic performance. The development of primordial engineering techniques by the Romans is largely due to the country's orography: 35 percent of the Italian territory is made up of mountains, 42 percent of hills, and 23 percent of plains.\footnote{ The Romans resorted to deviations in roads only when major obstacles could not be overcome by building structures such as bridges, and whenever possible road supports, like embankments or dykes, or tunnels through hills and mountains (\hyperlink{richard2010}{Richard, 2010}). Their roads in the Alps and the Apennines had steep slopes and allowed the movement of pedestrians, horses, and wagons.}

The investigation on the relationship between Roman roads and geography is first discussed using Figure \ref{figure6} where elevation data from Istat 
have been geo-coded and mapped using the polygon layer of the Italian territory. Then the layer of the Roman  network has been superimposed dividing roads into major (consular) and minor. The average altitude of each spatial unit has been classified according to five ordered equiproportional classes: [0 - 407 meters); [407 - 814); [814 - 1221);  [1221 - 1628); [1628 - 2035). The Italian map has been completed with the information on lakes and rivers using Corine Land Cover (CLC). 
The distribution of Roman roads among the different elevation classes is fairly homogeneous. The Roman infrastructure is also present in the darker areas, where elevation is higher.\footnote{ The right part of Figure \ref{figure6} zooms in on an exemplifying area of North-East Italy (i.e. the delimited rectangular area in the left part of the figure). The chosen area includes four different elevation zones, lakes and a stretch of Roman road that passes through lowlands and more elevated areas: the road does not circumnavigate the lake where the altitude is lower, but crosses a more elevated area.} In central-southern Italy, there is a high concentration of Roman roads in the Apennines, the second mountain range in Italy. Nevertheless, in the North, the highest concentration of Roman roads is along the Po Valley, where the average elevation is lower. Overall, the graphical evidence suggests that Romans approached the Italian territory in its complexity and variety, without leaving out any macro-area of the country in its actual borders without roads. Once it is clear that major/consular roads started from Rome and reached all the necessary outposts to expand the Empire outside the borders of our day Italy, it remains to ascertain what determined the position of the road in the mid-points. 

\begin{center}
[Figure \ref{figure6}]
\end{center}

Digging into the role of geography, the use of a thinner level of analysis provides detail on how the Romans approached the orography of the territory. 
The first purpose of constructing roads was to ease the movement of troops and vehicles (tanks) needed in military campaigns. Geography had a crucial role in making this task harder. The proximity of roads to other civil infrastructure or pre-existing settlements, instead, traced those economic factors that went beyond the need to move the Empire's borders. The list of information collected from distinct data sources and assembled for the following analysis is presented in Table \ref{table1}. 

\begin{center}
[Table \ref{table1}]
\end{center}

Both geographical and human activities features, which could affect the position of the network, are considered in the following model, where 10 x 10 \textit{km} grid cells are used as unit of observation to perform a conditional correlation test:

\begin{equation}\label{eq1}
\mathcal{RR}_{i} ={\alpha_{0} + \mathcal{\textbf{G}}_{i}  \alpha_{1} + \mathcal{\textbf{MA}}_{i} \alpha_{2}+ \phi_{p} + \textit{u}_{i}}
\end{equation}
where $\mathcal{RR}_{i}$ is the log-transformed measure of kilometres of major/consular roads;\footnote{ Both major and minor roads are tested in Section \ref{subsection_5.1}. } $\mathcal{\textbf{G}}_{i}$ denotes a matrix of several geographical measures, including the most common physical controls, as in  \hyperlink{dalgaard2018}{Dalgaard \textit{et al.} (2018)} and \hyperlink{flueckiger2022}{Flueckiger \textit{et al.} (2022)}. 
$\mathcal{\textbf{MA}}_{i}$ is a matrix that includes three different market access indicators;  $\phi_{p}$ are NUTS3 provincial fixed effects and $\textit{u}_{i}$ denotes the error term.

Among the geographical measures, \texttt{Ruggedness}$_{i}$, \texttt{Elevation}$_{i}$, \texttt{Slope}$_{i}$  capture the terrain's physical harshness, height and inclination. Terrain characteristics  also refer to their agriculture value or end-of-use: \texttt{Pre-1500 agriculture suitability}$_{i}$ measures the average cropland suitability as maximum calories yield before 1500. \texttt{Forest}$_{i}$ is a dummy variable equal to 1 if the grid cell is covered by forests or wooded vegetation.\footnote{ The information, sourced from \hyperlink{goldewijk2010}{Goldewijk (2010), ISLSCP II Historical Land Cover and Land Use, 1700-1990}, refers to the year 1700.} \texttt{Low vegetation}$_{i}$ is also a dummy variable and takes the value of 1 if the cell is covered by pastures or scrubland. Other geographical features, which are usually considered, refer to the proximity to water. \texttt{Distance from sea}$_{i}$ and \texttt{Distance from nearest waterway}$_{i}$  measure proximity to the nearest seacoast and river, respectively. They also capture those geographical elements related to market access and moving goods. Direct market access measures include, instead, the \texttt{Distance from nearest harbor}$_{i}$ and \texttt{Distance from Rome}$_{i}$, which capture, instead, the proximity to a harbor or port and the fact of moving out from Rome. Location details on settlements' presence have been included by assembling the information from  \hyperlink{bagnall2016}{Pleiades: A Gazetteer of Past Places}. \texttt{Pre-Roman amenities}$_i$ is a dummy variable that takes the value of 1 if in the unit of analysis existed simple pre-Roman settlements or settlements represented by a civil infrastructure (amphitheatre, theatre, cemetery, sanctuary, bath, bridges, ports, forts) before Romans.\footnote{ For  details see \url{https://pleiades.stoa.org/help/data-structure}.} 

The 10x10 \textit{km} grid level co-variance analysis is presented in Table \ref{table2}, where the effect of both geography and market access is measured on Roman roads, considering - in a first step - all 5,111 grid cells.\footnote{ The selection of the sole grid cells with Roman roads is accounted for in Table \ref{table3}.} Table \ref{table2} reveals that when only the orography of the territory is taken into account (Model (1) and (3)) the main geographical variable which is significantly correlated with major Roman roads is ruggedness, with an elasticity coefficient of -0.043 and -0.041, respectively. The first coefficient is not significantly different from the latter of Specification 3, where controls have been included to account for the territorial heterogeneity at NUTS3 level. The \texttt{Elevation$_{i}$} and \texttt{Slope}$_{i}$ measures add detail when a distinction is made across types of territory (Model (4), (5), (6)): mountain, hill or plain, respectively.\footnote{ Cells are classified according to three terrain zones: mountainous (if elevation is equal to or more than 700 meters); hilly (if elevation is less than 700 meters, but equal to or more than 300 meters); plain (if elevation is less than 300 meters).} On the one hand, the elasticity of -0.515 related to elevation discloses fewer kilometers of roads in challenging territories: the smaller value for hilly lands and the non significant coefficient for plain areas, instead, corroborates the reduced importance of the height in flatter cells. On the other hand, the -1.355  slope elasticity confirms that Roman roads are sparser in more impervious areas. 

Out of the different geographical features, agriculture suitability has been an attractor, mainly in hill terrains, while the presence of low vegetation is a deterrent. Proximity to sea and rivers - usually found as associated with settlements' birth and growth (\hyperlink{buringh2012}{Buringh \textit{et al.}, 2012}; \hyperlink{bosker2017}{Bosker and Buringh, 2017}) - do not play a role here. A predictable result is that grid cells of smaller size, composing the Italian coastlines, host less roads. 

\begin{center}
[Table \ref{table2}]
\end{center}

All in all, the results confirm the evidence in Figure \ref{figure6} on how the construction of Roman roads is linked to geography but not limited by it: roads are present in all types of terrains.

Integrating the territory's geography with market access measures (Model (2) and (7)), such as the presence of a harbor or the fact that all roads had Rome as a starting point, it emerges how both variables were not so relevant in explaining the Roman road path. Pre-Roman urbanization and settlements test whether Roman roads were built near larger urban centers or favored the subsequent expansion of earlier settlements. Existing cities could become logistic bases for organizing troops and military camps, providing infrastructure sound in war campaigns. In this design set, the position of existing settlements and cities may have contributed to determining the trajectory of a road segment or its terminal point, which varied by construction stage. Also, the road may have generated agglomeration effects leading to the subsequent development of cities. \texttt{Pre-Roman amenities} (any pre-Roman settlements, type of civil infrastructure or amenity already present) are positively correlated with longer roads.

Given the possible selection effect, since not all 5,111 grid cells have roads, Table 3 restricts the sample to the 1,359 grid cells that host major Roman roads. The role of ruggedness and elevation is absent. When focusing on the intra-grid variance (Model (3) and (4)), grid cells with higher slopes, instead, have fewer kilometers of roads. Results on the quality of the land and market access variables are all confirmed, with a weaker role played by the harbor distance.

\begin{center}
[Table \ref{table3}]
\end{center}

If geographical features determined - but not constrained - where Romans placed sections of road, once a grid cell is selected to host a road, the role of geography vanishes and more urban and market access factors become relevant in explaining the length of the network. This paper's first result will be used for investigating the link between old and new infrastructure in the following section. 

\section{\label{section_5} Old and modern infrastructure}

The legacy left by the historical Roman road network on current transport infrastructure - railways and motorways - is investigated by considering the possible factors behind their linkage.\footnote{ For an analysis on the intersection between the Roman road network and the Italian railway and motorway system see \nameref{AppendixB}.}

Cultural and landscape conditions determine why the new infrastructure may be related to the ancient one. The fact of having favored the birth and development of economic centers is an important one.\footnote{ Several contributions (see \hyperlink{garcia2015}{Garcia-Lopez \textit{et al.} (2015)} among others) stress that motorways are not located at random and argue in favor of the location of cities as the main driver of modern road infrastructure.} On the other hand, city location does not follow a casual process and has been found linked to (among other things) the presence of road infrastructure (\hyperlink{bosker2017}{Bosker and Buringh, 2017}). Besides, the direct effect that geography imposes on the costs of transport infrastructure projects cannot be overlooked. When landform creates construction difficulties, the presence of an old infrastructure facilitates the construction of a new one.

Detailed units of analysis allow taking into account how the modern infrastructure has been designed at the national level with an integrated view of the whole country's geography. The needs expressed by the local administrative authorities (regions and provinces) have been considered using controls at the NUTS3 level. Moreover, provinces also allow the highest within-country variance and pinpoint control for those historical legacies that followed the Roman Empire's collapse.\footnote{ An interesting strand of the literature focuses on the role of the social capital in Italy, exploiting the heterogeneity that originated from the events that followed the collapse of the Roman Empire (\hyperlink{guiso2004}{Guiso \textit{et al.}, 2004}).}
 
In Equation \eqref{eq2}, measures of modern infrastructures are regressed on the measure of Roman roads.
\vspace{-0.3cm}
\begin{equation}\label{eq2}
\mathcal{I}_{i} ={\beta_{0} +  \beta_{1} \mathcal{RR}_{i} +  \mathcal{\textbf{G}}_{i} \beta_{2} + \mathcal{\textbf{MA}}_{i} \beta_{3}  + \phi_{p} + \textit{u}_{i}}
\end{equation}

The dependent variable $\mathcal{I}_{i}$, is the log transformation of two measures in kilometers of current infrastructure: railways and motorways.\footnote{ Data on kilometers of railways at the grid level have been computed using Diva-GIS. Data on the current road network are from OpenStreetMap and are updated to 2009.} Considering two different transport systems derives from how transport infrastructure developed in Italy: railways from 1839, motorways from 1924 on.\footnote{ See \nameref{AppendixB}.} 
$\mathcal{\textbf{G}}_{i}$ is the matrix of the above discussed geographical measures, with agriculture suitability referring now to post-1500;
$\mathcal{\textbf{MA}}_{i}$ is the matrix of market access variables. The measure of \texttt{Pre-Roman amenities}$_{i}$ has been replaced by the binary variable \texttt{Large municipality}$_{i}$. It accounts for the urbanization of Italian cities prior the construction of the modern infrastructure, mimicking the role of pre-Roman settlements, civil infrastructures or amenities had in controlling for urban development before the expansion of Roman roads. The dummy variable takes the value of 1 if the grid cell $i$ had at least one municipality with more than 5,000 inhabitants between 1300 and 1861, before the new transport network was built.\footnote{ Data come from \hyperlink{malanima2015}{Malanima (2015)}: municipalities are considered as cities if their population was over 5,000 inhabitants.} The measure intends to capture the recursive effect from the old infrastructure to the birth and growth of cities and then back to the new transport infrastructure provision. NUTS3 provincial fixed effects $\phi_p$ complete the model specification, and $\textit{u}_{i}$ denotes the error term.

Table \ref{table4} presents the results of the model applied to the units of analysis (grid cells), containing segments of railways (Specifications from (1) to (4)) or motorways (Specifications from (5) to (8)). Regardless of the measure adopted, major (consular) roads are positively correlated with current infrastructure, slightly more in the case of motorways: the elasticity of $\mathcal{I}_{i}$ to $\mathcal{RR}_{i}$ is 0.30 - 0.34 for railways (without and with provincial controls, respectively) and 0.36 - 0.39 for motorways.  The effect of geography (with or without provincial controls) is confined to only some variables: \textit{Distance from sea}  in the case of railways and \textit{Low vegetation}  for motorways. When including market access measures, the development of urban municipalities before the unity of the Italian State is associated with a higher quantity of current infrastructure, with a coefficient of 0.48 and 0.39 for railways and motorways, respectively. Railways and motorways serve the connectivity of bigger cities, as extensively reported in the literature, where the fact of being connected to a major Roman road or a Roman hub makes the difference when accounting for the possibility of the smaller cities becoming larger centers (\hyperlink{bosker2013}{Bosker \textit{et al.}, 2013}).\footnote{ This validates the choice of including urban centers located nearby a major (consular) Roman road.}. The founding and growth of cities, expressing the need for new infrastructure, is at the heart of several recent contributions (\hyperlink{bosker2013}{Bosker \textit{et al.}, 2013}; \hyperlink{bosker2017}{Bosker and Buringh, 2017}; \hyperlink{michaels2018}{Michaels and Rauch, 2018}) on cities' geographical location. The reported results point to an independent effect of pre-existing urban development and old infrastructure on the current one, with coefficients on both variables large and consistently significant.

\begin{center}
[Table \ref{table4}]
\end{center}

To provide a control for multicollinearty and evaluate the importance of predictors,  Table \ref{table5} reports the output from running a Shapley-value regression analysis on Specifications (3), (4), (7) and (8) of Table \ref{table4}.  Results clearly show the 3- and 5-times larger role of Roman roads compared to urban municipalities in explaining the quantity of modern railways and motorways, respectively. Both variables contribute  more than 80\% to the $R^2$. All other geographical factors, instead, are relegated to a residual capability of explaining the current transport network.

\begin{center}
[Table \ref{table5}]
\end{center}

Estimations for all Italian grid cells - with and without railways or motorways - are available in Table \ref{tableA1} of Appendix \nameref{AppendixA}. Results are consistent with the evidence reported above about the predominance of Roman roads in determining the presence of infrastructure and how their legacy represented an important starting point in plain zones.

\subsection{\label{subsection_5.1} Robustness checks}

Robustness checks aim to assess the relevance of the Roman road index by considering variations in the definition of the old infrastructure measure and the set of territorial units included in the analysis. 

The first set of checks consists of the complete old infrastructure, a denser network including minor (non-consular) roads. If the starting node of the consular road network was concentrated in Rome, this was not the case for the entire system of Roman roads. The Romans constructed minor roads intending to enhance the connectedness of major roads by linking pre-Roman settlements or amenities. Their function of connecting consular roads by short routes or to urban centers made, in several cases, secondary roads as important as the major ones (\hyperlink{filiasi}{Filiasi, 1792}). Including minor roads in the analysis increases the kilometers of roads considered, re-scaling the $\mathcal{RR}_{i}$ variable used in Equation \ref{eq2} and enlarging the number of grid cells for determining the effect of geography and market access on the old infrastructure, as in Equation \ref{eq1}. Results in table \ref{table6} (columns 1-4) suggest that when also minor Roman roads are considered, the positive correlation between old and current infrastructure holds for both railways and motorways, with slightly larger coefficients in both equations (0.32 and 0.39, respectively) compared to the sole major roads specifications. Accounting for minor roads (Columns 5-6) refines the role of geography: now, a steeper terrain (slope) is associated with more kilometers of the old infrastructure. The construction of minor roads was less dependent from coastal layouts, did not concentrate on lands with higher agriculture values and was less distant from the harbors. The role of Rome, as the central node of the network is amplified, and the positive association with the presence of pre-Roman amenities is confirmed. Results are robust to any control included in the regression and the goodness-of-fit confirms the appropriateness of the estimated model.

\begin{center}
[Table \ref{table6}]
\end{center}

Table \ref{table7} returns to a measure of only major (consular) Roman roads now orthogonal to both geography and market access measures ($\tilde{\mathcal{RR}}_{i}$), i.e. the residuals of Equation \eqref{eq1} as in Model (4) of Table \ref{table3}. Including in the regression analysis only the component of the variance not explained by geography and market access leaves an index that allows measuring the role of Roman roads on the modern infrastructure independently from the effect that geographical and urbanization variables have on the old infrastructure. Estimates' precision is checked for possible spatial effects within 150 \textit{km} using Conley standard errors, as in Specifications 1 and 3, and bootstrap standard errors as in Specifications 2 and 4.\footnote{ As an estimator, the generated measure of Roman Roads has additional sampling variance that needs to be taken into account when calculating the variance of final parameter estimates. Bootstrap standard errors control for that.} Results are consistent with those presented in Section \ref{section_4}: the landform, captured by ruggedness, keeps a role in deterring the quantity of motorways, less in the case of railways. The size of the other effects (old infrastructure and large municipalities) are comparable with the results discussed above.

\begin{center}
[Table \ref{table7}]
\end{center}

The last set of robustness checks deals with the fact that all consular roads have a common starting point, Rome, determining a concentration of the old infrastructure in the territories nearby the capital. Specifications in Table \ref{table8} are obtained by excluding all grid cells for which the distance from Rome is less than 250 \textit{km}. Estimates have been replicated for both models (Equations \ref{eq1} and \ref{eq2}). Focusing on the old infrastructure (Columns 5 and 6), the slope coefficient is now more than double: it accounts for the fact that when moving from Rome, the Italian landform changes, including hills and mountains. The more challenging geography that did not impede the historical infrastructure construction.  Interestingly, results on the coefficients of the main determinants of interest do not change, while the variable \textit{Distance from Rome } looses all its explicative power. The discussion is similar when looking at the results of Equation \ref{eq2} (Specifications 1-4): the old infrastructure and the presence of a municipality determine the quantity of both railways and motorways. The size of the coefficients in both dependent outcomes is robust to the various controls.

\begin{center}
[Table \ref{table8}]
\end{center}

\section{\label{section_6}     Concluding remarks}

This paper provides novel evidence on the long-term effect of the Roman Empire. It shows how the Roman road network has had a persistent impact on the present-day road and railway system in Italy and how this represents one of the mechanisms of influence on modern economy.

The case of Italy is interesting for several reasons. Historically the Italian territory was characterized by a duality between the developed North-Center and a less-developed South. This duality persists today, and the differences between northern and southern Italy are still a matter. The central government is responsible for the institutions' functioning and the maintenance of the current infrastructure, but the Italian geographical decomposition is still reflected in road and railways development. The history of Italy after the fall of the Roman Empire was a collection of several dominations whose influence differed in different parts of the country. However, although the Roman Empire extended in the entire peninsula, the Roman road network showed remarkable variability: the result of the needs of the military campaigns conducted by the Romans during the Empire's expansion.

This paper focused on the consular Roman road network. Constructed for conquering new lands and with a linear shape between strategic points, they offer a valid exogenous index to measure the effect of the historical infrastructure on current motorways and railways.

Using detailed data at the grid cell level, the legacy of Roman roads on the modern infrastructure has been investigated in two steps. First, the role of confounding variables - both geographical  and more human and urban development characteristics - has been observed to understand the effect of these factors on the construction of Roman roads. The inferential results reveal how geography affected the building of the consular road network without limiting it. The Romans overcame physical barriers, and more demanding areas have not been excluded from developing an intricate system of connected roads. The analysis has been complemented by a descriptive investigation based on historical sources and geo-spatial maps, showing how the path of Roman roads was strictly driven by a military and straightness concept. Once a trajectory was conceived, Roman routes conformed more to a linear shape than to the most geographical advantageous layout.

In the second step, the effect of Roman roads on current railways and motorways has been tested. The econometric analysis displays a significant positive impact of the integrated ancient Roman road network on existing infrastructure. . Consular roads have been essential in determining the motorway network: this complements the qualitative information provided by the fact that, still today, some motorways in Italy are named after the Roman road that was constructed nearby. Results are robust to several controls aiming at testing the variety of Roman roads and the territory under scrutiny.

Regarding the mechanism driving the result, areas with a denser Roman road infrastructure are more likely to have a denser road and railway infrastructure today since Roman roads shaped the Italian landscape making the construction of current roads less costly. Today's roads simply followed the same path traced by the Romans. Reasons for this are quite diverse: in \hyperlink{bosker2017}{Bosker and Buringh (2017)} Roman roads are a proxy for favorable land-based accessibility locations and an important determinant for the founding of a city. In \hyperlink{bosker2013}{Bosker \textit{et al.} (2013)} Roman roads are taken as elements favoring subsequent urban expansion, differently from current roads, built for connecting already existing urban centers. Also, \hyperlink{michaels2018}{Michaels and Rauch (2018)} discuss how the presence of the old Roman road infrastructure contributes to path dependency in historical urban structuring. 

\newpage
\label{references}
\setcounter{secnumdepth}{0}
\section{References}

\hypertarget{acemoglu2001}{Acemoglu, D., S. Johnson and J.A. Robinson (2001), ``The Colonial Origins of Comparative Development: An Empirical Investigation,"  \textit{The American Economic Review}, 91(5): 1369-1401}.\\[-3.5ex]

\hspace{-0.6cm}\hypertarget{bagnall2016}{Bagnall, R. et al. (eds.) (2016), ``Pleiades: A Gazetteer of Past Places,'' \url{http://pleiades.stoa.org/} [Accessed: November 01, 2018]}.\\[-3.5ex]

\hspace{-0.6cm}\hypertarget{banerjee2012}{Banerjee, A., E. Duflo and N. Qian (2012), ``On the Road: Access to Transportation Infrastructure and Economic. Growth in China," NBER Working Paper N. 17897}.\\[-3.5ex] 


\hspace{-0.6cm}\hypertarget{ber2003}{Berechman, J. (2003), ``Transportation - Economic Aspects of Roman Highway Development: the Case of Via Appia,'' \textit{Transportation Research Part A: Policy and Practice}, 37(5): 453-478}. \\[-3.5ex] 

\hspace{-0.6cm}\hypertarget{berger2017}{Berger, T. and K. Enflo (2017), ``Locomotives of Local Growth: The Short- and Long-Term Impact of Railways in Sweden,'' \textit{Journal of Urban Economics}, 98(C): 124-138}.\\[-3.5ex] 

\hspace{-0.6cm} \hypertarget{bishop2014}{Bishop, M.C. (2014), \textit{The Secret History of the Roman Roads of Britain}, Pen \& Sword Military}.\\[-3.5ex] 

\hspace{-0.6cm}\hypertarget{bosker2017}{Bosker, M. and E. Buringh (2017), ``City Seeds: Geography and the Origins of the European City System,'' \textit{Journal of Urban Economics}, 98: 139-157}.\\[-3.5ex] 

\hspace{-0.6cm}\hypertarget{bosker2013}{Bosker, M., E. Buringh and J.L. van Zanden (2013), ``From Baghdad to London: Unraveling Urban Development in Europe, the Middle East, and North Africa, 800-1800,'' \textit{The Review of Economics and Statistics}, 95(4): 1418-1437}.\\[-3.5ex]

\hspace{-0.6cm} \hypertarget{brooks2009}{Brooks, D.H. and D. Hummels (2009), \textit{Infrastructure's Role in Lowering Asia's Trade Costs: Building for Trade}, Edward Elgar Publishing}.\\[-3.5ex]

\hspace{-0.6cm}\hypertarget{buringh2012}{Buringh, E., J.L van Zanden and M. Bosker (2012), ``Soldiers and Booze: The Rise and Decline of a Roman Market Economy in North-Western Europe,'' Utrecht University, CGEH Working Paper No. 32}.\\[-3.5ex] 

\hspace{-0.6cm}\hypertarget{carla2017}{Carl\`a-Uhink, F. (2017), \textit{The ``Birth" of Italy. The Institutionalization of Italy as a Region, 3rd-1st Century BCE}, Berlin and Boston: De Gruyter}.\\[-3.5ex]

\hspace{-0.6cm}\hypertarget{chevallier1976}{Chevallier, R. (1976), \textit{Roman Roads}, trans. N.H. Field, Berkeley: University of California press, 1976}.\\[-3.5ex]

\hspace{-0.6cm}\hypertarget{cornell1995}{Cornell, T. (1995), \textit{The Beginnings of Rome: Italy and Rome from the Bronze Age to the Punic Wars}, Routledge}.\\[-3.5ex]

\hspace{-0.6cm} \hypertarget{cornell1982}{Cornell, T. and J. Matthews (1982), \textit{Atlas of the Roman World}, New York, N.Y.: Facts on File}.\\[-3.5ex]
 
\hspace{-0.6cm}\hypertarget{crescenzi2016}{Crescenzi, R., Di Cataldo, M and A. Rodr\'iguez-Pose (2016), ``Government Quality and the Economic Returns of Transport Infrastructure Investment in European Regions,"  \textit{Journal of Regional Science}, 56(4): 555-582}.\\[-3.5ex]

\hspace{-0.6cm} \hypertarget{dalgaard2018}{Dalgaard, C.J., N. Kaarsen, O. Olsson and P. Selaya (2018), ``Roman Roads to Prosperity: Persistence and Non-Persistence of Public Goods Provision,'' CEPR Discussion Paper 12745}.\\[-3.5ex]
 
\hspace{-0.6cm}\hypertarget{degraauw2014}{de Graauw, A., B. Maione-Downing and M. McCormick (2013), ``Geodatabase of Ancient Ports and Harbors", DARMC Scholarly Data Series, Data Contribution Series 2013-2}.\\[-3.5ex]


\hspace{-0.6cm}\hypertarget{donaldson2018}{Donaldson, D. (2018), ``Railways of the Raj: Estimating the Impact of Transportation Infrastructure,'' \textit{The American Economic Review}, 108(4-5): 899-934}.\\[-3.5ex]

\hspace{-0.6cm}\hypertarget{donaldson2016}{Donaldson, D. and R. Hornbeck (2016), ``Railways and American Economic Growth: A `Market Access' Approach,'' \textit{The Quarterly Journal of Economics}, 131(2): 799-858}.\\[-3.5ex]

\hspace{-0.6cm}\hypertarget{faber2014}{Faber, B. (2014), ``Trade Integration, Market Size, and Industrialization: Evidence from China's National Trunk Highway System,'' \textit{The Review of Economic Studies}, 81(3): 1046-1070}.\\[-3.5ex]

\hspace{-0.6cm}\hypertarget{faigelbaum2014}{Fajgelbaum, P. and S.J. Redding (2014), ``External Integration, Structural Transformation and Economic Development: Evidence from Argentina 1870-1914,'' NBER Working Paper No.
20217}.\\[-3.5ex]

\hspace{-0.7cm} \hypertarget{filiasi1792}{Filiasi, G. (1792), \textit{Delle Strade Romane che Passavano Anticamente pel Mantovano}, Arnaldo Forni Editore}.\\[-3.5ex]

\hspace{-0.7cm} \hypertarget{finley1973}{Finley, M.I. (1973), \textit{The Ancient Economy}, Berkeley and Los Angeles: University of California Press}.\\[-3.5ex]

\hspace{-0.6cm}\hypertarget{flueckiger2022}{Flueckiger, M., E. Hornung, M. Larch, M. Ludwig and A. Mees (2022), ``Roman Transport Network Connectivity and Economic Integration,'' \textit{Review of Economic Studies}, 89: 774-810}.\\[-3.5ex]

\hspace{-0.7cm} \hypertarget{fosbroke1843}{Fosbroke, T.D. (1843), \textit{Encyclopaedia of Antiquities, and Elements of Archaeology, Classical and Mediaeval Volume 2}, M.A. Nattali, London}.\\[-3.5ex] 
 
\hspace{-0.6cm}\hypertarget{galor2016}{Galor, O. and  \"{O}. \"{O}zak (2016), ``The Agricultural Origins of Time Preference,'' \textit{American Economic Review}, 106(10): 3064-3103}.\\[-3.5ex]

\hspace{-0.6cm}\hypertarget{garcia2015}{Garcia-L\'{o}pez, M.\`{A}., A. Holl and E. Viladecans-Marsal (2015), ``Suburbanization and Highways in Spain when the Romans and the Bourbons still Shape its Cities'', \textit{Journal of Urban Economics}, 85: 52-67}.\\[-3.5ex]

\hspace{-0.6cm}\hypertarget{gleason2013}{Gleason, J.P. (2013), \textit{Roman Roads in Gaul: How Lines of Communication and Basing Support Operational Reach}, School of Advanced Military Studies Monographs}.\\[-3.5ex]
 
\hspace{-0.6cm}\hypertarget{goldewijk2010}{Goldewijk, K.K. (2010), ``ISLSCP II Historical Land Cover and Land Use, 1700-1990'', in Hall, Forest G., G. Collatz, B. Meeson, S. Los, E. Brown de Colstoun, and D. Landis (eds.), ISLSCP Initiative II Collection, \url{http://daac.ornl.gov/} [Accessed: March 07, 2022]}.\\[-3.5ex]

\hspace{-0.6cm}\hypertarget{guiso2004}{Guiso, L., P. Sapienza and L. Zingales (2004), ``The Role of Social Capital in Financial Development,'' \textit{The American Economic Review,} 94(3): 526-556}.\\[-3.5ex]

\hspace{-0.6cm}\hypertarget{hindle1998}{Hindle, P. (1998), \textit{Medieval Roads and Tracks}, ed.3, Princes Risborough}.\\[-3.5ex]
 
\hspace{-0.6cm}\hypertarget{jarvis2008}{Jarvis, A., H.I. Reuter, A.  Nelson and E. Guevara (2008), ``Hole-filled seamless SRTM data V4,'' International Centre for Tropical  Agriculture (CIAT), \url{http://srtm.csi.cgiar.org} [Accessed: April 02, 2020]}.\\[-3.5ex]

\hspace{-0.6cm}\hypertarget{jedwab2017}{Jedwab, R., E. Kerby and A. Moradi (2017), ``History, Path Dependence and Development: Evidence from Colonial Railways, Settlers and Cities in Kenya,'' \textit{The Economic Journal}, 127 (603): 1467-1494}.\\[-3.5ex]

 \hspace{-0.6cm}\hypertarget{laporta1997}{La Porta, R., F. Lopez-de-Silanes, A. Shleifer and R.W. Vishny (1997), ``Legal  Determinants of External Finance,'' \textit{Journal of Finance}, 52(3): 1131-1150}.\\[-3.5ex]
 
\hspace{-0.6cm}\hypertarget{laurence1999}{Laurence, R. (1999), \textit{The Roads of Roman Italy: Mobility and Cultural Change}, Routledge, London}.\\[-3.5ex]

\hspace{-0.6cm}\hypertarget{licio2021}{Licio, V. (2021), ``When History Leaves a Mark: A New Measure of Roman Roads,'' \textit{Italian Economic Journal}, 7(1): 1-35}.\\[-3.5ex]

\hspace{-0.6cm}\hypertarget{lopez1956}{Lopez, R.S. (1956), ``The Evolution of Land Transport in the Middle Ages,'' \textit{Past and Present}, 9(1): 17-29}.\\[-3.5ex]

\hspace{-0.6cm}\hypertarget{malanima2005}{Malanima, P. (2005), ``Urbanisation and the Italian Economy during the Last Millennium,'' \textit{European Review of Economic History}, 9(1): 97-122}.\\[-3.5ex]

\hspace{-0.6cm}\hypertarget{malanima2015}{Malanima, P. (2015), ``Italian Urban Population 1300-1861 (The Database),'' 2005-revision 2015, \url{http://www.paolomalanima.it/default_file/Italian\%20Economy/Urban_Population01.pdf} [Accessed: April 06, 2020]}.\\[-3.5ex]

\hspace{-0.6cm}\hypertarget{margary1973}{Margary, I.D. (1973), \textit{Roman Roads in Britain}, John Baker Publishing, London}.\\[-3.5ex]

\hspace{-0.6cm}\hypertarget{mccormick2013}{McCormick, M., G. Huang, G. Zambotti and J. Lavash (2013), ``Roman Road Network (version 2008),'' DARMC Scholarly Data Series, Data Contribution Series 2013-5}.\\[-3.5ex]

\hspace{-0.6cm}\hypertarget{michalopoulos2017}{Michalopoulos, S. and E. Papaioannou (2017),  \textit{The Long Economic and Political Shadow of History}, Vol. I, II and III, Cepr Press, London}.\\[-3.5ex]

\hspace{-0.6cm}\hypertarget{michaels2018}{Michaels, G. and F. Rauch (2018), ``Resetting the Urban Network: 117-2012,'' \textit{The Economic Journal}, 128(608): 378-412}.\\[-3.5ex]

\hspace{-0.6cm}\hypertarget{musti1990}{Musti, D. (1990),  \textit{La Spinta Verso il Sud: Espansione Romana e Rapporti ``Internazionali"}, in Arnaldo Momigliano; Aldo Schiavone (edited by), Storia di Roma Vol. I, Torino, Einaudi}.\\[-3.5ex]

\hspace{-0.6cm}\hypertarget{nunn2009}{Nunn, N. (2009), ``The Importance of History for Economic Development,'' \textit{Annual Review of Econonomics}, 1(1): 65-92}.\\[-3.5ex]
 
\hspace{-0.6cm}\hypertarget{nunn2012}{Nunn, N. and D. Puga (2012), ``Ruggedness: The blessing of bad geography in Africa'', \textit{Review of Economics and Statistics}, 94(1), 20-36}.\\[-3.5ex]

\hspace{-0.6cm} \hypertarget{percoco2013}{Percoco, M. (2013), ``Geography, Institutions and Urban Development: Italian Cities, 1300-1861,'' \textit{The Annals of Regional Science}, 50: 135-152}.\\[-3.5ex]

\hspace{-0.6cm}\hypertarget{poulter2010}{Poulter, J. (2010), \textit{The Planning of Roman Roads and Walls in Northern Britain}, Amberley Publishing}.\\[-3.5ex]

\hspace{-0.6cm}\hypertarget{ramcharan2009}{Ramcharan, R. (2009), ``Why an Economic Core: Domestic Transport Costs?,'' \textit{Journal of Economic Geography}, 9(4): 559-581}.\\[-3.5ex]

\hspace{-0.6cm}\hypertarget{richard2010}{Richard, C.J. (2010), \textit{Why We're All Romans: The Roman Contribution to the Western World}, Rowman \& Littlefield Publishers}.\\[-3.5ex]

\hspace{-0.6cm}\hypertarget{roberts2020}{Roberts, M., M. Melecky, T. Bougna and Y.S. Xu (2020), ``Transport Corridors and their Wider Economic Benefits: A Quantitative Review of the Literature", \textit{Journal of Regional Science}, 60(2): 207-248}.\\[-3.5ex]

\hspace{-0.6cm}\hypertarget{smith1890}{Smith, W. (1890), \textit{A Dictionary of Greek and Roman Antiquities}, London: John Murray, 1890}.\\[-3.5ex]

\hspace{-0.6cm}\hypertarget{temin2013}{Temin, P. (2013), \textit{The Roman Market Economy}, Princeton: Princeton University Press}.\\[-3.5ex]

\hspace{-0.6cm}\hypertarget{talbert2000}{Talbert, R.J. (ed.) (2000), \textit{Barrington Atlas of the Greek and Roman World}, Princeton: Princeton University Press}.\\[-3.5ex]

\hspace{-0.6cm}\hypertarget{volpe2014}{Volpe Martincus C., J. Carballo and A. Cusolito (2014), ``Routes, Exports, and Employment in Developing Countries: Following the Trace of the Inca Roads," mimeo}.\\[-3.5ex]

\hspace{-0.6cm}\hypertarget{vonhagen1967}{Von Hagen, V.W. (1967), \textit{The Roads That Led to Rome}, The World Publishing Company, Cleveland and New York}.\\[-3.5ex]

\hspace{-0.6cm}\hypertarget{wahl2017}{Wahl, F. (2017), ``Does European Development have Roman Roots? Evidence from the German Limes," \textit{Journal of Economic Growth}, 22(3): 313-349}.\\[-3.5ex]

\hspace{-0.6cm}\hypertarget{welfare1995}{Welfare, H. and V. Swan (1995), \textit{Roman Camps in England: The Field Archaeology}, London: HMSO}.

\newpage
\setcounter{secnumdepth}{0}
\section{Figures and tables}
\renewcommand{\thetable}{\arabic{table}}
\setcounter{table}{0}  
\renewcommand{\thefigure}{\arabic{figure}}
\renewcommand{\theHfigure}{\arabic{figure}}
\setcounter{figure}{0}  

\begin{figure}[h]
\begin{center}
\caption{\label{figure1} Roman road layer and Italian 10 x 10 \textit{km} grid cells}
\includegraphics[scale=0.55]{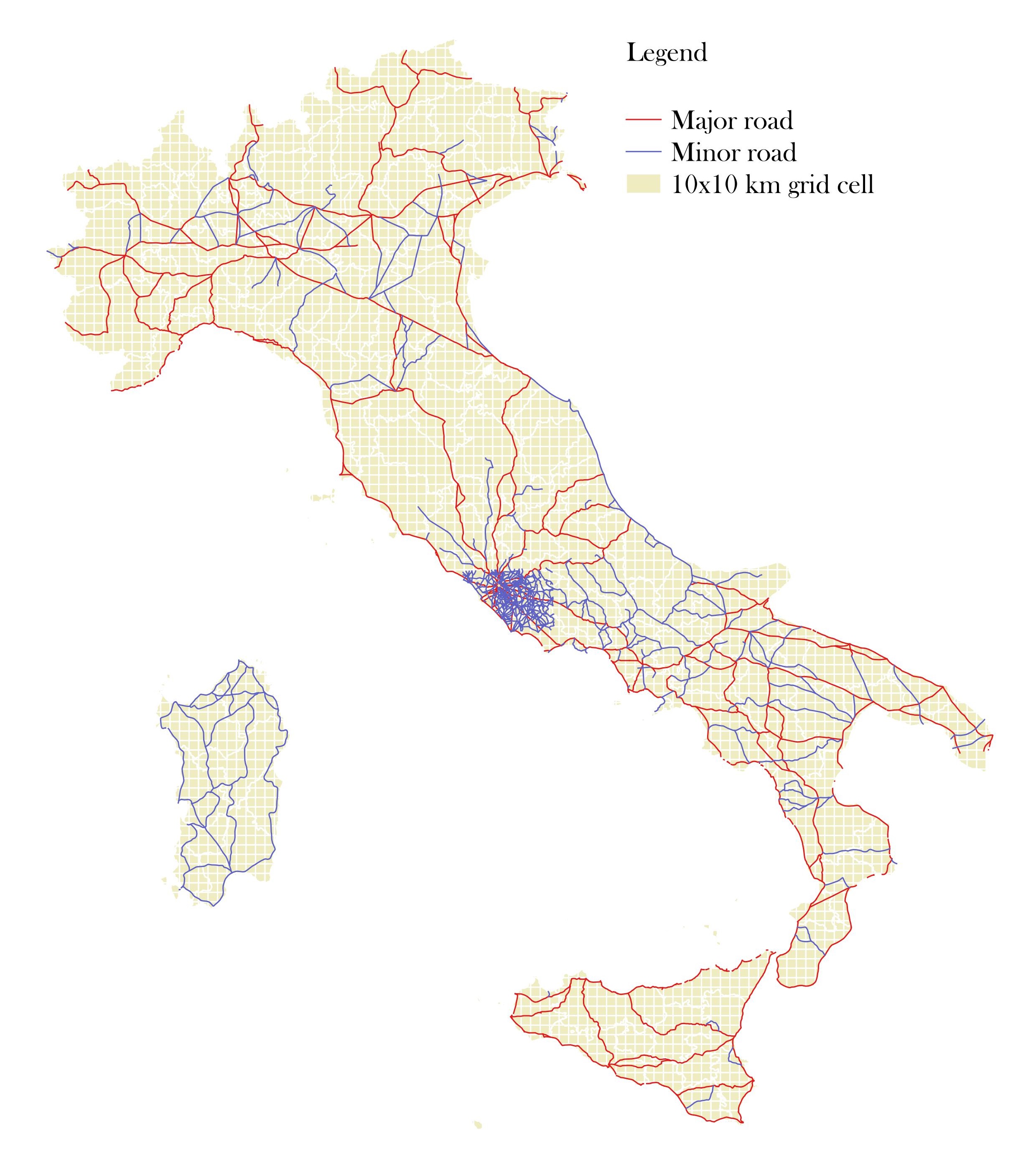}
\end{center}
\footnotesize{Source: Authors' elaboration from McCormick, M. \textit{et al.} 2013. ``Roman Road Network (version 2008),'' DARMC Scholarly Data Series 2013-5 and from Istat data (2011)}
\end{figure}

\newpage
\begin{figure}[h]
\begin{center}
\caption{\label{figure2} Roman Empire and Via Appia in 312 B.C.}
\includegraphics[scale=0.7]{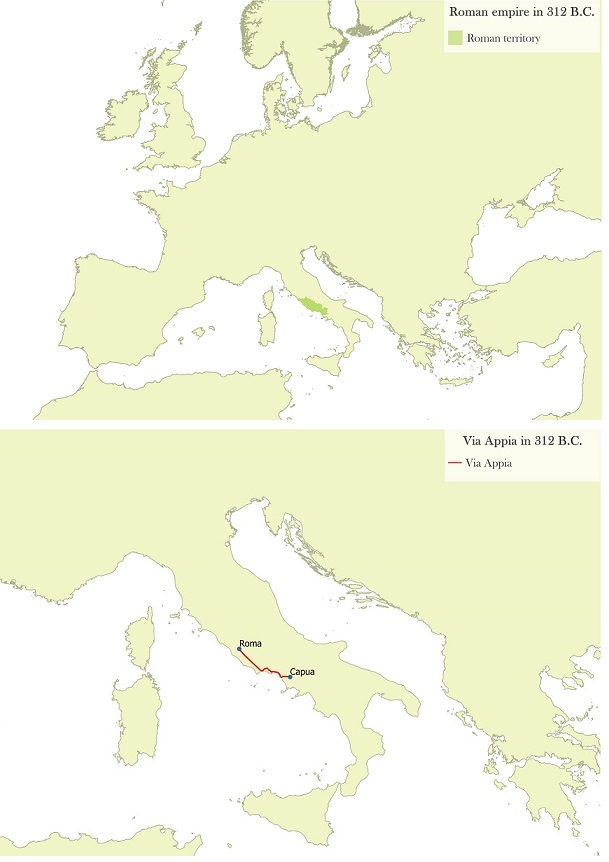}
\end{center}
\vspace{-0.6cm}
\begin{center}
\footnotesize{Source: Authors' drawing}
\end{center}
\end{figure}

\newpage
\begin{figure}[h] 
\begin{center}
\caption{\label{figure3} Roman Empire and Via Appia in 238 B.C.}
\includegraphics[scale=0.7]{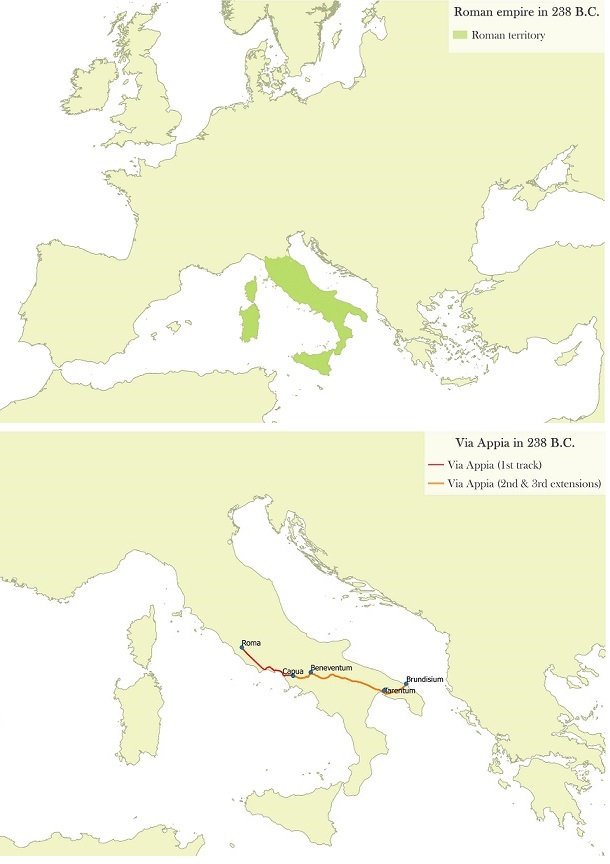}
\end{center}
\vspace{-0.6cm}
\begin{center}
\footnotesize{Source: Authors' drawing}
\end{center}
\end{figure}

\newpage
\begin{figure}[h] 
\begin{center}
\caption{\label{figure4} Romans' expansionist objectives and the conquest of Greece}
\includegraphics[scale=0.7]{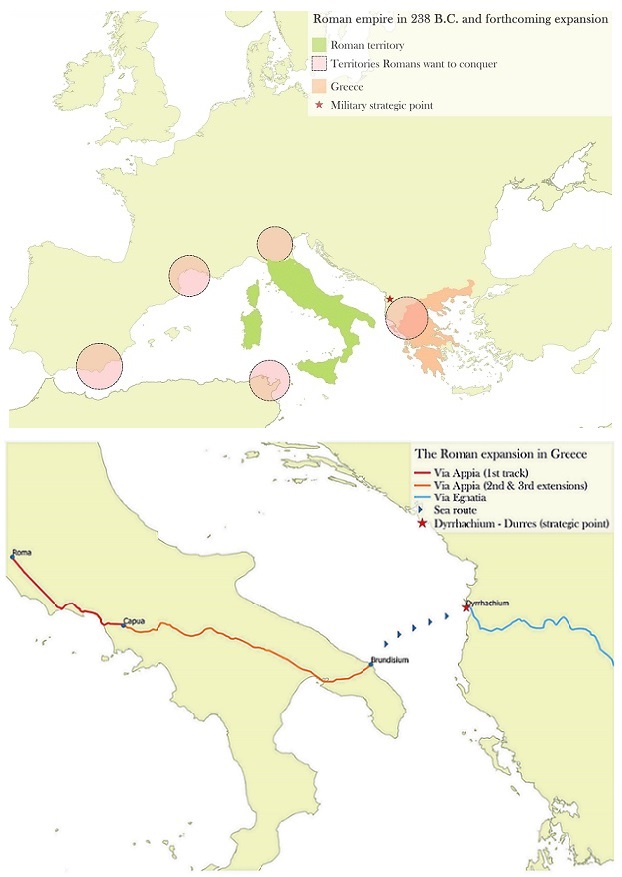}
\end{center}
\vspace{-0.6cm}
\begin{center}
\footnotesize{Source: Authors' drawing}
\end{center}
\end{figure}

\newpage
\begin{figure}[h] 
\begin{center}
\caption{\label{figure5} The Via Annia}
\includegraphics[scale=0.50]{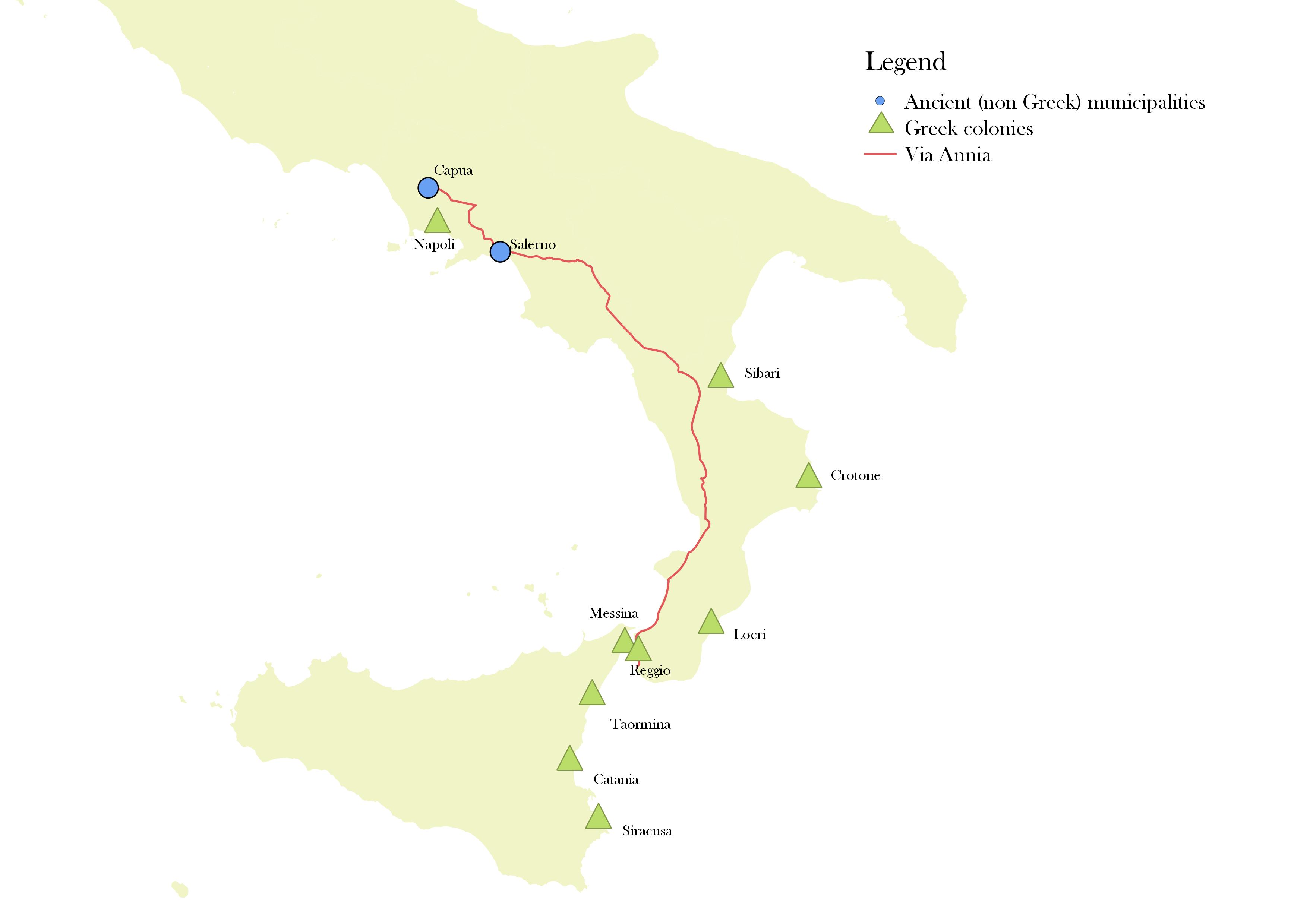}
\end{center}
\vspace{-0.6cm}
\begin{center}
\footnotesize{Source: Authors' drawing}
\end{center}
\end{figure}

\newpage
\begin{figure}[h]
\begin{center}
\caption{\label{figure6} Roman roads and geography: position by elevation}
\includegraphics[scale=0.13]{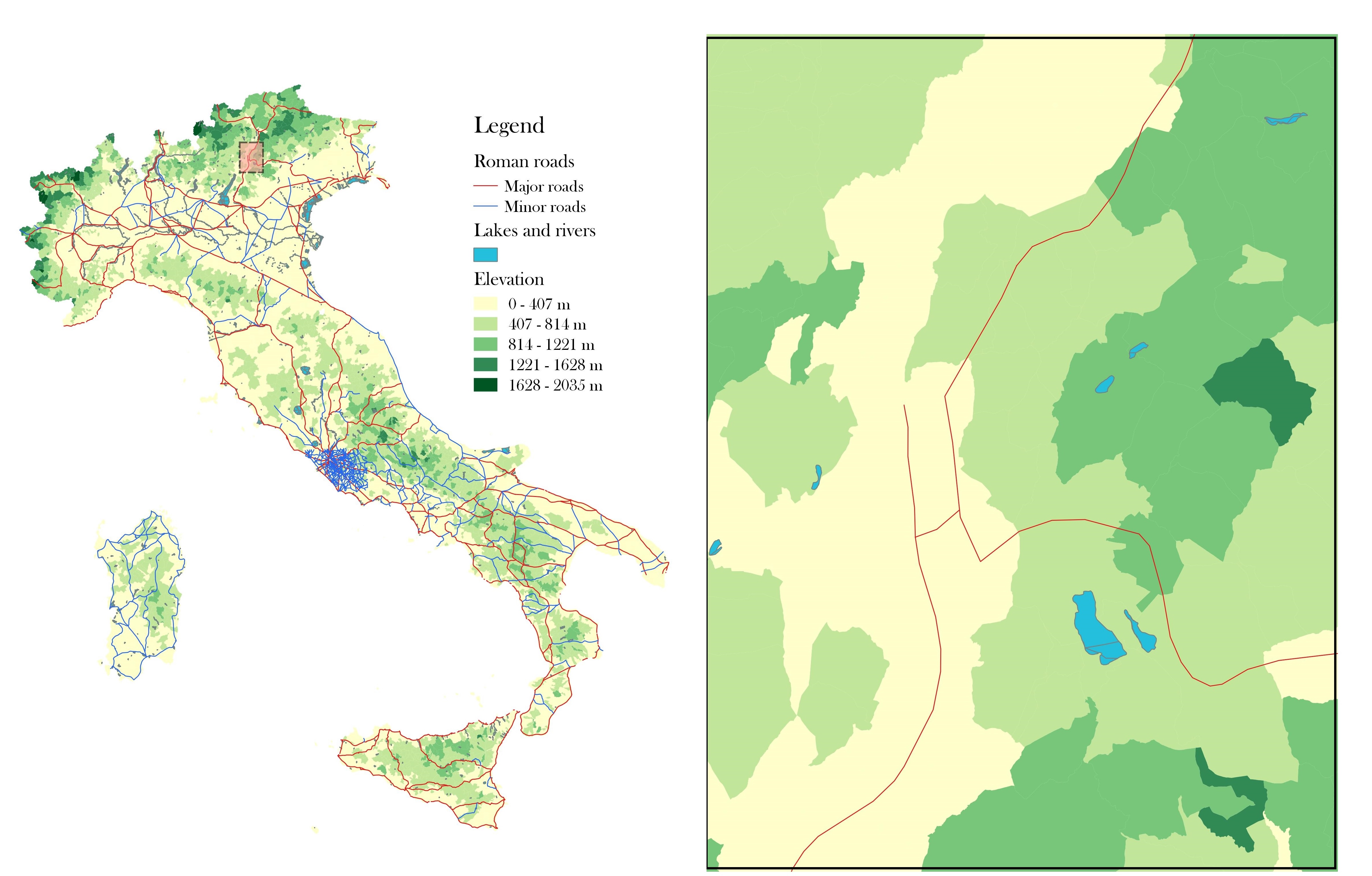}
\end{center}
\footnotesize{Source: Authors' drawing from Istat data, Corine Land Cover data, McCormick, M. \textit{et al.} 2013. ``Roman Road Network (version 2008),'' DARMC Scholarly Data Series 2013-5}
\end{figure}

\newpage
\begin{landscape}
\begin{table} [h!]
\caption{\label{table1} Variables and sources}
\vspace{-0.6cm}
\begin{center}\tiny
\begin{tabular}{l l l l l}
\hline\hline\\
\textit{Variables} & \textit{Definition} & \textit{Years} & \textit{Source} & \textit{Available for}\\
\hline\\
Major Roman roads&	Length of major Roman roads in the grid cell (km)&	117  &	\hyperlink{licio2021}{Licio's (2021)} elaboration from \hyperlink{mccormick2013}{McCormick \textit{et al.} (2013)}&	5,111 10x10 km cells\\
\\
All Roman roads&	Length of all Roman roads in the grid cell (km)&	117  &	\hyperlink{licio2021}{Licio's (2021)} elaboration from \hyperlink{mccormick2013}{McCormick \textit{et al.} (2013)} &	5,111 10x10 km cells\\
\\
Major Roman roads (residuals) & Length of major Roman roads in the grid cell (km)&	117 &	Authors' elaboration  &	5,111 10x10 km cells\\[-0.5ex]
&partialling out geography and market access&&&\\
\\
Railways& Length of current railways in the grid cell (km)&	Modern &	Authors' elaboration from Diva-GIS  &	5,111 10x10 km cells\\
\\
Motorways& Length of current motorways in the grid cell (km)&	2009 &	Authors' elaboration from OpenStreetMap  &	5,111 10x10 km cells\\
\\
Mountain  &	Dummy variable: 1 if the mean elevation of the&	Time invariant&	Authors' elaboration from \hyperlink{jarvis2008}{Jarvis \textit{et al.} (2008)} ``Hole-filled  seamless    &	5,104 10x10 km cells \\[-0.5ex]
& grid cell is $\geq 700$ m&&SRTM data V4'' , International  Centre for Tropical  Agriculture (CIAT)& \\
\\
Hill  &	Dummy variable: 1 if the mean elevation of the&	Time invariant&	Authors' elaboration from \hyperlink{jarvis2008}{Jarvis \textit{et al.} (2008)} ``Hole-filled  seamless    &	5,104 10x10 km cells \\[-0.5ex]
&grid cell is $\geq$300 m, but $<$700 m&&SRTM  data V4'' , International  Centre for Tropical  Agriculture (CIAT)& \\
\\
Plain  &	Dummy variable: 1 if the mean elevation of the&	Time invariant&	Authors' elaboration from \hyperlink{jarvis2008}{Jarvis \textit{et al.} (2008)}``Hole-filled  seamless  &	5,104 10x10 km cells \\[-0.5ex]
&grid cell is $<$300 m&&SRTM  data V4'' , International  Centre for Tropical  Agriculture (CIAT)& \\
\\
Elevation  &	Average terrain elevation of the grid cell (m) &	Time invariant&	Authors' elaboration from \hyperlink{jarvis2008}{Jarvis \textit{et al.} (2008)} ``Hole-filled  seamless   &	5,104 10x10 km cells \\[-0.5ex]
&&&SRTM  data V4'' , International  Centre for Tropical  Agriculture (CIAT)& \\
\\
Ruggedness &Average of the Terrain Ruggedness Index in the &Time invariant&Authors' elaboration from \hyperlink{nunn2012}{Nunn and Puga (2012)}&5,111 10x10 km cells   \\
& grid cell (hundred m)&&&\\
\\
Slope  &	Average terrain slope in the grid cell (degrees)&	Time invariant&	Authors' elaboration from \hyperlink{jarvis2008}{Jarvis \textit{et al.} (2008)} ``Hole-filled  seamless   &	5,095 10x10 km cells \\[-0.5ex]
&&&SRTM  data V4'' , International  Centre for Tropical  Agriculture (CIAT)& \\
\\
 Pre 1500 cropland suitability& Average of cropland suitability, as maximum &	Pre 1500&	Authors' elaboration from \hyperlink{galor2016}{Galor and \"{O}zak (2016)}    &	5,111 10x10 km cells \\
&calories yield, in the grid cell &&&   \\
 \\
Post 1500 cropland suitability& Average of cropland suitability, as maximum&	Post 1500&	Authors' elaboration from \hyperlink{galor2016}{Galor and \"{O}zak (2016)}   &	5,111 10x10 km cells \\
 & calories yield, in the grid cell &&&   \\
 \\
Forest &Dummy variable: 1 if the grid cell was covered&1700&Authors' elaboration from \hyperlink{goldewijk2010}{Goldewijk (2010)} ``ISLSCP II Historical  &5,111 10x10 km cells   \\
 & by forests or wooded vegetation &&Land Cover and Land Use, 1700-1990''&   \\
\\
Low vegetation &Dummy variable: 1 if the grid cell was covered&1700&Authors' elaboration from \hyperlink{goldewijk2010}{Goldewijk (2010)} ``ISLSCP II Historical &5,111 10x10 km cells   \\
 & by pastures or scrub lands &&Land Cover and Land Use, 1700-1990''&   \\
\\
Distance from sea &Distance from the nearest seacoast (km)  &Time invariant&Authors' elaboration&5,111 10x10 km cells   \\
&to the grid cell's centroid&&&\\
\\
Distance from nearest waterway &Distance from nearest river (km) &Time invariant&Authors' elaboration from OpenStreetMap&5,111 10x10 km cells   \\
&to the grid cell's centroid&&&\\
\\
Smaller area &Dummy variable: 1 if the area of the grid cell&Time invariant&Authors' elaboration&5,111 10x10 km cells   \\
 & is $<$ 100 \textit{km}$^{2}$  (cells in the borders)&&&   \\
 \\
Distance from nearest harbor &Distance from nearest ancient harbor or port (km)&Time invariant&Authors' elaboration from \hyperlink{degraauw2014}{de Graauw \textit{et al.} (2013)} ``Geodatabase &5,111 10x10 km cells   \\
&to the grid cell's centroid&&of Ancient Ports and Harbors''&\\
\\
Distance from Rome &Distance from the city of Rome (km)&Time invariant&Authors' elaboration&5,111 10x10 km cells   \\
&to the grid cell's centroid&&&\\
\\
Pre-Roman amenities &	Dummy variable: 1 if in the grid cell existed&	Before 30 B.C.&	Authors' elaboration from \hyperlink{bagnall2016}{Pleiades}&	5,111 10x10 km cells\\
&settlements, civil infrastructures or amenities &&&\\
&(amphitheaters, theaters, cemeteries, sanctuaries,&&&\\
&baths, bridges, ports, forts) before the Romans&&&\\
\\
Large municipality &	Dummy variable: 1 if the grid cell had at least  &	1300-1861 &Authors' elaboration from \hyperlink{malanima2015}{Malanima (2015)}&	5,111 10x10 km cells\\
& 1 municipality with more than 5,000 inhabitants&&&\\
& during the 5 centuries (1300-1861) before the &&&\\
& construction of the current transport infrastructure &&&\\
\hline \hline
\end{tabular}
\end{center}
\vspace{-0.4cm}
\scriptsize{Source: Authors' elaborations}
\end{table} 
\end{landscape}

\begin{landscape}
\begin{table}[h]
\caption{\label{table2} Determinants of constructing major Roman roads (all cells)}
\vspace{-0.7cm}
\begin{center}\footnotesize
\begin{threeparttable}
\begin{tabular}{l c c c  c c c c}
\hline\hline\\
 & \textit{(1)} & \textit{(2)} & \textit{(3)} & \textit{(4)} & \textit{(5)} & \textit{(6)} & \textit{(7)}\\[1ex]
 \textit{Dependent variable:}&&&&&&&\\[1ex]
\textit{Major Roman roads in km (log)}&Total&Total&Total&Mountain&Hill&Plain&Total \\[1ex]
\hline\\
\texttt{Ruggedness} (log)&      -0.043** &      -0.045** &      -0.041** &       0.084   &      -0.047   &      -0.063*  &      -0.045** \\
                    &     (0.019)   &     (0.018)   &     (0.019)   &     (0.082)   &     (0.094)   &     (0.037)   &     (0.018)   \\
\texttt{Elevation} (log)&      -0.019   &      -0.036*  &      -0.105***&      -0.515***&      -0.243***&       0.006   &      -0.092***\\
                    &     (0.030)   &     (0.019)   &     (0.010)   &     (0.084)   &     (0.054)   &     (0.039)   &     (0.012)   \\
\texttt{Slope} (log)&       0.077*  &       0.040   &       0.096** &      -1.355*  &       0.156   &       0.080*  &       0.078** \\
                    &     (0.042)   &     (0.030)   &     (0.041)   &     (0.732)   &     (0.151)   &     (0.041)   &     (0.037)   \\
\texttt{Agriculture suitability pre 1500} (log) &       0.018** &       0.003   &       0.041***&       0.019** &       0.117***&       0.016   &       0.031***\\
                    &     (0.007)   &     (0.013)   &     (0.008)   &     (0.008)   &     (0.032)   &     (0.023)   &     (0.006)   \\
\texttt{Forest} (dummy)&       0.027   &       0.037   &       0.067   &       0.250***&       0.114   &      -0.042   &       0.078   \\
                    &     (0.064)   &     (0.033)   &     (0.071)   &     (0.031)   &     (0.177)   &     (0.049)   &     (0.069)   \\
\texttt{Low vegetation} (dummy)&      -0.086*  &      -0.102***&      -0.073** &       0.095***&       0.012   &      -0.075***&      -0.063*  \\
                    &     (0.050)   &     (0.038)   &     (0.034)   &     (0.035)   &     (0.094)   &     (0.016)   &     (0.034)   \\
\texttt{Distance from sea} (log)  &      -0.033   &       0.016   &       0.029   &       0.096   &      -0.110   &       0.052*  &       0.036*  \\
                    &     (0.034)   &     (0.027)   &     (0.025)   &     (0.098)   &     (0.106)   &     (0.028)   &     (0.022)   \\
\texttt{Distance from nearest waterway} (log) &       0.019   &      -0.013   &      -0.023   &      -0.004   &      -0.044   &      -0.016   &      -0.016   \\
                    &     (0.018)   &     (0.017)   &     (0.023)   &     (0.036)   &     (0.033)   &     (0.042)   &     (0.020)   \\
\texttt{Distance from nearest harbor} (log)&               &      -0.105** &               &               &               &               &      -0.114***\\
                    &               &     (0.043)   &               &               &               &               &     (0.040)   \\
\texttt{Distance from Rome} (log)&               &       0.049   &               &               &               &               &      -0.435** \\
                    &               &     (0.109)   &               &               &               &               &     (0.190)   \\
\texttt{Pre-Roman amenities} (dummy)&               &       0.684***&               &               &               &               &       0.638***\\
                    &               &     (0.078)   &               &               &               &               &     (0.076)   \\
\hline\\
 $ \phi_{p}$	&No&No&Yes&Yes&Yes&Yes&Yes \\  
 \texttt{Smaller area} (dummy) &Yes&Yes&Yes&Yes&Yes&Yes&Yes \\ 
Observations  &        5,095&        5,095&        5,095&        1,371&        1,304&        2,420&        5,095\\
Adjusted \textit{R$^{2}$}&       0.201&       0.267&       0.269&       0.196&       0.253&       0.336&       0.324\\
\hline \hline
\end{tabular}
\begin{tablenotes}[para,flushleft]
\scriptsize
\item Note: All log-transformed variables are indicated with (log).  $ \phi_{p}$ represents NUTS3 province fixed effects. Conley standard errors are reported in parentheses and calculated with a spatial cutoff of 150 kilometers. Results are confirmed for shorter and longer distances.
Asterisks denote significance levels; * p$<$0.10, ** p$<$0.05 and *** p$<$0.01. 
\end{tablenotes}
\end{threeparttable}
\end{center}
\end{table} 
\end{landscape}

\begin{table}[h]
\caption{\label{table3} Determinants of major Roman roads (non-zero cells)}
\vspace{-0.7cm}
\begin{center}\footnotesize
\begin{threeparttable}
\begin{tabular}{l c c c c }
\hline\hline\\
& \textit{(1)} & \textit{(2)} & \textit{(3)} & \textit{(4)}\\[1ex]
\textit{Dependent variable:} &&&& \\[1ex]
\textit{Major Roman roads in km (log)}&&&& \\[1ex]
\hline\\
\texttt{Ruggedness} (log)&      -0.016   &      -0.037   &      -0.052   &      -0.056   \\
                    &     (0.033)   &     (0.036)   &     (0.057)   &     (0.060)   \\
\texttt{Elevation} (log)&       0.015   &       0.014   &       0.036   &       0.037   \\
                    &     (0.047)   &     (0.046)   &     (0.071)   &     (0.071)   \\
\texttt{Slope} (log)&       0.056   &      -0.005   &      -0.226** &      -0.207*  \\
                    &     (0.117)   &     (0.132)   &     (0.109)   &     (0.121)   \\
\texttt{Agriculture suitability pre 1500} (log) &       0.031***&       0.029   &       0.065***&       0.066***\\
                    &     (0.010)   &     (0.021)   &     (0.022)   &     (0.023)   \\
\texttt{Forest} (dummy)&      -0.069   &      -0.077   &       0.030   &       0.037   \\
                    &     (0.063)   &         (0.071)   &     (0.057)   &     (0.067)   \\
\texttt{Low vegetation} (dummy)&      -0.035   &      -0.040   &      -0.015   &      -0.030   \\
                    &     (0.046)   &     (0.072)   &     (0.063)   &     (0.081)   \\
\texttt{Distance from sea} (log)  &       0.031   &       0.055***&       0.089***&       0.101** \\
                    &     (0.021)   &     (0.020)   &     (0.032)   &     (0.039)   \\
\texttt{Distance from nearest waterway} (log) &       0.015   &      -0.005   &      -0.020   &      -0.003   \\
                    &     (0.025)   &     (0.023)   &     (0.039)   &     (0.033)   \\
\texttt{Distance from nearest harbor} (log)&               &      -0.038   &               &      -0.103*  \\
                    &               &     (0.037)   &               &     (0.054)   \\
\texttt{Distance from Rome} (log)&               &       0.018   &               &      -0.419** \\
                    &               &     (0.046)   &               &     (0.180)   \\
\texttt{Pre-Roman amenities} (dummy)&               &       0.500***&               &       0.515***\\
                    &               &     (0.091)   &               &     (0.102)   \\
  \hline\\
 $ \phi_{p}$	&No&No&Yes&Yes\\  
 \texttt{Smaller area} (dummy)&Yes&Yes&Yes&Yes\\  
 Observations&        1,359&        1,359&        1,359&        1,359\\
 Adjusted \textit{R$^{2}$}&       0.682&       0.695&       0.682&       0.697\\
\hline \hline           
\end{tabular}
\begin{tablenotes}[para,flushleft]
\scriptsize
\item Note: All log-transformed variables are indicated with (log).  $ \phi_{p}$ represents NUTS3 province fixed effects. Conley standard errors are reported in parentheses and calculated with a spatial cutoff of 150 kilometers. Results are confirmed for shorter and longer distances.
Asterisks denote significance levels; * p$<$0.10, ** p$<$0.05 and *** p$<$0.01.
\end{tablenotes}
\end{threeparttable}
\end{center}
\end{table}

\begin{landscape}
\begin{table}[h]
\caption{\label{table4} Determinants of modern transport infrastructure (non-zero cells)}
\vspace{-0.7cm}
\begin{center}\footnotesize
\begin{threeparttable}
\begin{tabular}{l c c c c  c  c c c c }
\hline\hline\\
 & \textit{(1)} & \textit{(2)} & \textit{(3)} & \textit{(4)} & \textit{(5)} & \textit{(6)} & \textit{(7)}& \textit{(8)}\\[1ex]
\textit{Dependent variable:}\\[1ex]
\textit{Current infrastructure in km (log)}&\multicolumn{4}{c}{\textit{Railways}}&\multicolumn{4}{c}{\textit{Motorways}}\\[1ex]
\hline\\
\texttt{Major} $\mathcal{RR}_{i} $ (log)&       0.301***&       0.273***&       0.336***&       0.298***&       0.362***&       0.340***&       0.393***&       0.362***\\
                    &     (0.023)   &     (0.023)   &     (0.022)   &     (0.022)   &     (0.020)   &     (0.012)   &     (0.029)   &     (0.030)   \\
\texttt{Ruggedness} (log)&      -0.054   &      -0.041   &      -0.067   &      -0.048   &      -0.123** &      -0.096*  &      -0.163*  &      -0.122   \\
                    &     (0.045)   &     (0.050)   &     (0.046)   &     (0.047)   &     (0.056)   &     (0.058)   &     (0.090)   &     (0.103)   \\
\texttt{Elevation} (log)&       0.060   &       0.047   &       0.071   &       0.080   &       0.093   &       0.077   &       0.100   &       0.076   \\
                    &     (0.046)   &     (0.053)   &     (0.062)   &     (0.057)   &     (0.083)   &     (0.089)   &     (0.137)   &     (0.142)   \\
\texttt{Slope} (log)&      -0.146   &      -0.329   &      -0.110   &      -0.290   &      -0.045   &      -0.322   &      -0.107   &      -0.370   \\
                    &     (0.274)   &     (0.274)   &     (0.307)   &     (0.322)   &     (0.654)   &     (0.643)   &     (0.470)   &     (0.521)   \\
\texttt{Agriculture suitability post 1500} (log) &       0.019   &       0.004   &       0.000   &      -0.014   &      -0.002   &      -0.006   &      -0.016   &      -0.029   \\
                    &     (0.015)   &     (0.018)   &     (0.027)   &     (0.028)   &     (0.027)   &     (0.025)   &     (0.032)   &     (0.030)   \\
\texttt{Forest} (dummy)&      -0.083   &      -0.052   &      -0.156*  &      -0.110   &      -0.091   &      -0.053   &      -0.225   &      -0.217   \\
                    &     (0.067)   &     (0.075)   &     (0.088)   &     (0.082)   &     (0.078)   &     (0.109)   &     (0.165)   &     (0.142)   \\
\texttt{Low vegetation} (dummy)&       0.040   &       0.034   &       0.031   &       0.057   &      -0.181***&      -0.175** &      -0.187***&      -0.131*  \\
                    &     (0.035)   &     (0.045)   &     (0.036)   &     (0.036)   &     (0.063)   &     (0.085)   &     (0.067)   &     (0.074)   \\
\texttt{Distance from sea} (log)  &       0.058***&       0.072***&       0.116** &       0.093** &       0.006   &       0.009   &       0.126   &       0.157   \\
                    &     (0.003)   &     (0.025)   &     (0.046)   &     (0.045)   &     (0.040)   &     (0.042)   &     (0.111)   &     (0.113)   \\
\texttt{Distance from nearest waterway} (log) &       0.023   &       0.001   &       0.022   &       0.028   &       0.032   &       0.027   &       0.049   &       0.044   \\
                    &     (0.018)   &     (0.022)   &     (0.030)   &     (0.029)   &     (0.034)   &     (0.036)   &     (0.057)   &     (0.058)   \\
\texttt{Distance from nearest harbor} (log)&               &      -0.008   &               &      -0.058   &               &       0.037   &               &      -0.078   \\
                    &               &     (0.039)   &               &     (0.070)   &               &     (0.035)   &               &     (0.064)   \\
\texttt{Distance from Rome} (log)&               &       0.002   &               &      -0.136   &               &       0.019   &               &       0.421***\\
                    &               &     (0.035)   &               &     (0.142)   &               &     (0.050)   &               &     (0.075)   \\
\texttt{Large municipality} (dummy)&               &       0.352***&               &       0.479***&               &       0.305***&               &       0.386***\\
                    &               &     (0.040)   &               &     (0.053)   &               &     (0.080)   &               &     (0.112)   \\
\hline\\
 $ \phi_{p}$	&No&No&Yes&Yes&No&No&Yes&Yes\\ 
  \texttt{Smaller area} (dummy)&Yes&Yes&Yes&Yes&Yes&Yes&Yes&Yes\\
Observations&  2,236&        2,236&        2,236&        2,236&         907&         907&         907&         907\\
Adjusted \textit{R$^{2}$}&0.747&       0.753&       0.749&       0.758&       0.684&       0.688&       0.675&       0.682\\
\hline \hline           
\end{tabular}
\begin{tablenotes}[para,flushleft]
\scriptsize
\item Note: All log-transformed variables are indicated with (log).  $ \phi_{p}$ represents NUTS3 province fixed effects. Conley standard errors are reported in parentheses and calculated with a spatial cutoff of 150 kilometers. Results are confirmed for shorter and longer distances.
Asterisks denote significance levels; * p$<$0.10, ** p$<$0.05 and *** p$<$0.01.
\end{tablenotes}
\end{threeparttable}
\end{center}
\end{table} 
\end{landscape}

\begin{table}[h]
\caption{\label{table5} Shapley-value approach for determinants of modern transport infrastructure}
\vspace{-0.7cm}
\begin{center}\footnotesize
\begin{threeparttable}
\begin{tabular}{l c c c c }
\hline\hline\\
& \textit{(1)} & \textit{(2)} & \textit{(3)} & \textit{(4)}\\[1ex]
\\
&\multicolumn{4}{c}{ \textbf{\% Shapley-value}}\\[1ex]
\textit{Dependent variable:} &&&& \\[1ex]
\textit{Current infrastructure in km (log)}&\multicolumn{2}{c}{\textit{Railways}}&\multicolumn{2}{c}{\textit{Motorways}}\\[1ex]
\hline\\
\texttt{Major} $\mathcal{RR}_{i} $ (log)&80.26&60.52&83.72&69.22\\
\texttt{Ruggedness} (log)&6.67&5.08&3.97&3.20\\
\texttt{Elevation} (log)&3.32&2.65&1.96&1.62\\
\texttt{Slope} (log)&0.23&0.13&0.13&0.08\\
\texttt{Agriculture suitability post 1500} (log) &6.33&4.47&2.31&1.56\\
\texttt{Forest} (dummy)&1.13&0.67&0.83&0.70\\
\texttt{Low vegetation} (dummy)&0.15&0.17&2.06&2.11\\
\texttt{Distance from sea} (log) &1.66&2.06&0.50&0.61\\
\texttt{Distance from nearest waterway} (log)&0.25&0.32&4.52&5.66\\
\texttt{Smaller area} (dummy)&0.00&0.00&0.00&0.00\\
\texttt{Distance from nearest harbor} (log)&&0.72&&0.34\\
\texttt{Distance from Rome} (log)&&0.33&&0.11\\
\texttt{Large municipality} (dummy)&&22.86&&14.79\\
\hline \hline           
\end{tabular}
\begin{tablenotes}[para,flushleft]
\scriptsize
\item Note: Shapley decomposition refers to Specification (3), (4), (7), and (8) of Table \ref{table4}.
\end{tablenotes}
\end{threeparttable}
\end{center}
\end{table}

\begin{landscape}
\begin{table}[h]
\caption{\label{table6} Robustness checks: all Roman roads}
\vspace{-0.7cm}
\begin{center}\scriptsize
\begin{threeparttable}
\begin{tabular}{l c c c  c c c }
\hline\hline\\
 & \textit{(1)} & \textit{(2)} & \textit{(3)} & \textit{(4)} & \textit{(5)} & \textit{(6)}\\[1ex]
\textit{Dependent variable:} &\multicolumn{4}{c}{{}}&\multicolumn{2}{c}{{}}\\[1ex]
\textit{Transport infrastructure in km (log)}&\multicolumn{2}{c}{\textit{Railways}}&\multicolumn{2}{c}{\textit{Motorways}}&\multicolumn{2}{c}{\textit{All} $\mathcal{RR}$}\\[1ex]
\hline\\
\texttt{All} $\mathcal{RR}_{i} $ (log)&       0.299***&       0.323***&       0.357***&       0.385***&               &               \\
                    &     (0.020)   &     (0.018)   &     (0.028)   &     (0.038)   &               &               \\
\texttt{Ruggedness} (log)&      -0.028   &      -0.040   &      -0.088*  &      -0.125   &      -0.024   &       0.006   \\
                    &     (0.044)   &     (0.044)   &     (0.053)   &     (0.088)   &     (0.042)   &     (0.059)   \\
\texttt{Elevation} (log)&       0.035   &       0.078   &       0.066   &       0.069   &      -0.018   &      -0.076   \\
                    &     (0.046)   &     (0.051)   &     (0.084)   &     (0.129)   &     (0.051)   &     (0.073)   \\
\texttt{Slope} (log)&      -0.416   &      -0.411   &      -0.596   &      -0.600   &       0.426***&       0.574***\\
                    &     (0.287)   &     (0.353)   &     (0.699)   &     (0.606)   &     (0.144)   &     (0.133)   \\
\texttt{Agriculture suitability post 1500} (log) &       0.003   &      -0.019   &      -0.004   &      -0.020   &               &               \\
                    &     (0.015)   &     (0.023)   &     (0.030)   &     (0.032)   &               &               \\
\texttt{Forest} (dummy)&      -0.017   &      -0.077   &      -0.005   &      -0.171   &      -0.036   &       0.090   \\
                    &     (0.080)   &     (0.078)   &     (0.089)   &     (0.128)   &     (0.060)   &     (0.134)   \\
\texttt{Low vegetation} (dummy)&       0.020   &       0.070   &      -0.155   &      -0.090   &      -0.037   &      -0.079   \\
                    &     (0.048)   &     (0.054)   &     (0.101)   &     (0.077)   &     (0.057)   &     (0.083)   \\
\texttt{Distance from sea} (log)  &       0.065***&       0.088*  &       0.009   &       0.144   &       0.096***&       0.158***\\
                    &     (0.025)   &     (0.052)   &     (0.028)   &     (0.107)   &     (0.034)   &     (0.044)   \\
\texttt{Distance from nearest waterway} (log) &      -0.016   &       0.009   &       0.020   &       0.031   &       0.027   &       0.059*  \\
                    &     (0.025)   &     (0.029)   &     (0.034)   &     (0.058)   &     (0.021)   &     (0.034)   \\
\texttt{Distance from nearest harbor} (log)&      -0.011   &      -0.084   &       0.008   &      -0.090   &      -0.088*  &      -0.104***\\
                    &     (0.034)   &     (0.069)   &     (0.035)   &     (0.067)   &     (0.049)   &     (0.032)   \\
\texttt{Distance from Rome} (log)&       0.111** &      -0.020   &       0.132***&       0.479***&      -0.170***&      -0.648** \\
                    &     (0.047)   &     (0.121)   &     (0.035)   &     (0.075)   &     (0.056)   &     (0.264)   \\
\texttt{Large municipality} (dummy)&       0.329***&       0.416***&       0.265***&       0.335***&               &               \\
                    &     (0.026)   &     (0.054)   &     (0.080)   &     (0.111)   &               &               \\ 
\texttt{Agriculture suitability pre 1500} (log) &               &               &               &               &      -0.022   &       0.011   \\
                    &               &               &               &               &     (0.030)   &     (0.027)   \\
\texttt{Pre-Roman amenities} (dummy)&               &               &               &               &       0.495***&       0.550***\\
                    &               &               &               &               &     (0.041)   &     (0.041)   \\           
\hline\\
 $ \phi_{p}$	&No&Yes&No&Yes&No&Yes\\ 
   \texttt{Smaller area} (dummy)&Yes&Yes&Yes&Yes&Yes&Yes\\
Observations&        2,236&        2,236&         907&         907&        2,068&        2,068\\
Adjusted \textit{R$^{2}$}&       0.760&       0.766&       0.697&       0.693&              0.738&       0.746\\
\hline \hline   
\end{tabular}
\begin{tablenotes}[para,flushleft]
\scriptsize
\item Note: All log-transformed variables are indicated with (log).  $ \phi_{p}$ represents NUTS3 province fixed effects. Conley standard errors are reported in parentheses and calculated with a spatial cutoff of 150 kilometers. Results are confirmed for shorter and longer distances.
Asterisks denote significance levels; * p$<$0.10, ** p$<$0.05 and *** p$<$0.01.
\end{tablenotes}
\end{threeparttable}
\end{center}
\end{table} 
\end{landscape}

\begin{table}[h]
\caption{\label{table7} Robustness checks: major Roman roads as residuals of Model (4) of Table \ref{table3}}
\vspace{-0.7cm}
\begin{center}\footnotesize
\begin{threeparttable}
\begin{tabular}{l c c c c }
\hline\hline\\
& \textit{(1)} & \textit{(2)} & \textit{(3)} & \textit{(4)}\\[1ex]
\textit{Dependent variable:} &&&& \\[1ex]
\textit{Current infrastructure in km (log)}&\multicolumn{2}{c}{\textit{Railways}}&\multicolumn{2}{c}{\textit{Motorways}}\\[1ex]
\hline\\
\texttt{Major} $\tilde{\mathcal{RR}}_{i} $ (log)&       0.272***&       0.272***&       0.352***&       0.352***\\
                    &     (0.024)   &     (0.025)   &     (0.027)   &     (0.042)   \\
\texttt{Ruggedness} (log)&      -0.060   &      -0.060*  &      -0.132   &      -0.132** \\
                    &     (0.047)   &     (0.036)   &     (0.093)   &     (0.067)   \\
\texttt{Elevation} (log)&       0.090   &       0.090*  &       0.088   &       0.088   \\
                    &     (0.058)   &     (0.048)   &     (0.132)   &     (0.091)   \\
\texttt{Slope} (log)&      -0.369   &      -0.369   &      -0.468   &      -0.468   \\
                    &     (0.315)   &     (0.365)   &     (0.437)   &     (0.775)   \\
\texttt{Agriculture suitability post 1500} (log) &       0.005   &       0.005   &      -0.002   &      -0.002   \\
                    &     (0.031)   &     (0.029)   &     (0.031)   &     (0.052)   \\
\texttt{Forest} (dummy)&      -0.088   &      -0.088   &      -0.195   &      -0.195   \\
                    &     (0.081)   &     (0.094)   &     (0.162)   &     (0.168)   \\
\texttt{Low vegetation} (dummy)&       0.053   &       0.053   &      -0.128*  &      -0.128   \\
                    &     (0.041)   &     (0.074)   &     (0.077)   &     (0.147)   \\
\texttt{Distance from sea} (log)  &       0.116** &       0.116***&       0.186   &       0.186** \\
                    &     (0.046)   &     (0.043)   &     (0.117)   &     (0.095)   \\
\texttt{Distance from nearest waterway} (log) &       0.025   &       0.025   &       0.040   &       0.040   \\
                    &     (0.030)   &     (0.027)   &     (0.061)   &     (0.048)   \\
\texttt{Distance from nearest harbor} (log)&      -0.093   &      -0.093   &      -0.126*  &      -0.126   \\
                    &     (0.072)   &     (0.059)   &     (0.068)   &     (0.105)   \\
\texttt{Distance from Rome} (log)&      -0.265*  &      -0.265*  &       0.304   &       0.304   \\
                    &     (0.143)   &     (0.151)   &         (0.369)   &     (0.377)   \\
\texttt{Large municipality} (dummy)&       0.499***&       0.499***&       0.408***&       0.408***\\
                    &     (0.055)   &     (0.053)   &     (0.108)   &     (0.098)   \\
  \hline\\
 $ \phi_{p}$	&Yes&Yes&Yes&Yes\\  
 \texttt{Smaller area} (dummy)&Yes&Yes&Yes&Yes\\  
 Observations&        2,236&        2,236&        907&        907\\
 Adjusted \textit{R$^{2}$}&       0.754&       0.754&       0.677&       0.677\\
\hline \hline           
\end{tabular}
\begin{tablenotes}[para,flushleft]
\scriptsize
\item Note: All log-transformed variables are indicated with (log).  $ \phi_{p}$ represents NUTS3 province fixed effects. For Specification (1) and (3) Conley  standard errors are reported in parentheses and calculated with a spatial cutoff of 150 kilometers. Results are confirmed for shorter and longer distances. For Specification (2) and (4) bootstrap standard errors (10,000 replications) are reported in parentheses. Asterisks denote significance levels; * p$<$0.10, ** p$<$0.05 and *** p$<$0.01.
\end{tablenotes}
\end{threeparttable}
\end{center}
\end{table}

\begin{landscape}
\begin{table}[h]
\caption{\label{table8} Robustness checks: no cells surrounding Rome (distance $>$ 250 \textit{km})}
\vspace{-0.7cm}
\begin{center}\scriptsize
\begin{threeparttable}
\begin{tabular}{l c c c  c c c }
\hline\hline\\
 & \textit{(1)} & \textit{(2)} & \textit{(3)} & \textit{(4)} & \textit{(5)} & \textit{(6)}\\[1ex]
\textit{Dependent variable:} &\multicolumn{4}{c}{{}}&\multicolumn{2}{c}{{}}\\[1ex]
\textit{Transport infrastructure in km (log)}&\multicolumn{2}{c}{\textit{Railways}}&\multicolumn{2}{c}{\textit{Motorways}}&\multicolumn{2}{c}{\textit{Major} $\mathcal{RR}$}\\[1ex]
\hline\\
 \texttt{Major} $\mathcal{RR}_{i} $ (log)&       0.286***&       0.315***&       0.336***   &       0.365***&               &               \\
                    &     (0.022)   &     (0.016)   &         (0.026)   &     (0.024)   &               &               \\
\texttt{Ruggedness} (log)&      -0.024   &      -0.025   &      -0.051   &      -0.057   &       0.009   &       0.047   \\
                    &     (0.047)   &     (0.051)   &     (0.042)   &     (0.064)   &     (0.050)   &     (0.074)   \\
\texttt{Elevation} (log)&       0.025   &       0.053   &      -0.001   &       0.006   &      -0.023   &      -0.096   \\
                    &     (0.049)   &     (0.063)   &     (0.066)   &     (0.100)   &     (0.067)   &     (0.082)   \\
\texttt{Slope} (log)&      -0.269   &      -0.187   &       0.196   &       0.024   &       0.466***&       0.474***\\
                    &     (0.213)   &     (0.326)   &     (0.131)   &     (0.118)   &     (0.159)   &     (0.157)   \\
\texttt{Agriculture suitability post 1500} (log) &      -0.003   &      -0.014   &      -0.004   &      -0.029   &               &               \\
                    &     (0.026)   &     (0.029)   &     (0.027)   &     (0.040)   &               &               \\
\texttt{Forest} (dummy)&      -0.058   &      -0.089   &       0.004   &      -0.138   &      -0.026   &      -0.050   \\
                    &     (0.116)   &     (0.100)   &     (0.143)   &     (0.175)   &     (0.065)   &     (0.119)   \\
\texttt{Low vegetation} (dummy)&       0.033   &       0.032** &      -0.140   &      -0.048   &      -0.030   &      -0.098   \\
                    &     (0.028)   &     (0.015)   &     (0.099)   &         (0.023)   &     (0.051)   &     (0.091)   \\
\texttt{Distance from sea} (log)  &       0.082** &       0.125** &       0.051   &       0.209   &       0.099** &       0.188***\\
                    &     (0.032)   &     (0.050)   &     (0.050)   &     (0.148)   &     (0.046)   &     (0.053)   \\
\texttt{Distance from nearest waterway} (log) &      -0.003   &       0.048*  &       0.038   &       0.091   &       0.035   &       0.045   \\
                    &     (0.029)   &     (0.028)   &     (0.048)   &     (0.064)   &     (0.026)   &     (0.041)   \\
\texttt{Distance from nearest harbor} (log)&      -0.007   &      -0.078   &       0.018   &      -0.124   &      -0.039   &      -0.120** \\
                    &     (0.052)   &     (0.091)   &     (0.056)   &     (0.086)   &     (0.040)   &     (0.048)   \\
\texttt{Distance from Rome} (log)&      -0.136   &      -0.285   &       0.160   &       0.702   &       0.102   &       0.526   \\
                    &     (0.180)   &     (0.346)   &     (0.235)   &     (1.160)   &     (0.149)   &     (0.648)   \\
\texttt{Large municipality} (dummy)&       0.350***&       0.520***&       0.390***&       0.563***&               &               \\
                    &     (0.062)   &     (0.077)   &     (0.097)   &     (0.108)   &               &               \\       
 \texttt{Agriculture suitability pre 1500} (log) &               &               &               &               &       0.018   &       0.017   \\
                    &               &               &               &               &     (0.023)   &     (0.030)   \\
\texttt{Pre-Roman amenities} (dummy)&               &               &               &               &       0.479***&       0.507***\\
                    &               &               &               &               &     (0.060)   &     (0.060)   \\                    
\hline\\
 $ \phi_{p}$	&No&Yes&No&Yes&No&Yes\\ 
   \texttt{Smaller area} (dummy)&Yes&Yes&Yes&Yes&Yes&Yes\\
Observations&      1,669&       1,669&         657&         657&        1,489&        1,489\\
Adjusted \textit{R$^{2}$}&       0.756&       0.763&       0.690&       0.690&              0.737&       0.738\\
\hline \hline   
\end{tabular}
\begin{tablenotes}[para,flushleft]
\scriptsize
\item Note: All log-transformed variables are indicated with (log).  $ \phi_{p}$ represents NUTS3 province fixed effects. Conley standard errors are reported in parentheses and calculated with a spatial cutoff of 150 kilometers. Results are confirmed for shorter and longer distances.
Asterisks denote significance levels; * p$<$0.10, ** p$<$0.05 and *** p$<$0.01.
\end{tablenotes}
\end{threeparttable}
\end{center}
\end{table} 
\end{landscape}

\clearpage
\newpage
\setcounter{secnumdepth}{0}
\section{Appendix A - Additional figures and tables}
\label{AppendixA}
\appendix
\renewcommand{\thetable}{A.\arabic{table}}
\setcounter{table}{0}  
\renewcommand{\thefigure}{A.\arabic{figure}}
\renewcommand{\theHfigure}{A.\arabic{figure}}
\setcounter{figure}{0}  

\begin{figure}[h!]
\begin{center}
\caption{\label{figureA1} Main major Roman roads in Italy}
\includegraphics[scale=0.50]{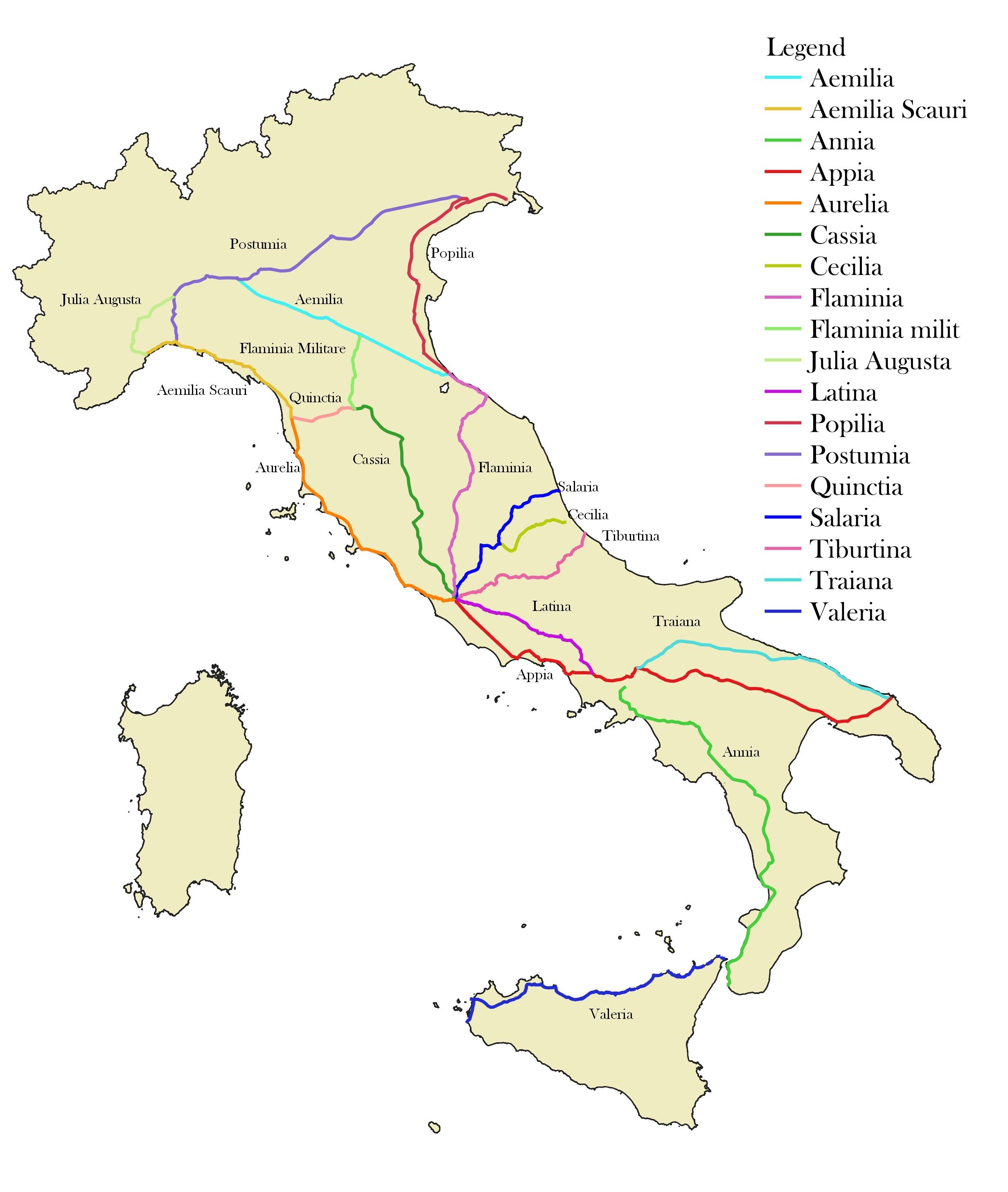}
\end{center}
\footnotesize{Source: Authors' drawing from McCormick, M. \textit{et al.} 2013. ``Roman Road Network (version 2008),'' DARMC Scholarly Data Series 2013-5  }
\end{figure}

\begin{figure}[h!]
\begin{center}
\caption{\label{figureA2} Roman roads by importance: major and minor roads}
\includegraphics[scale=0.50]{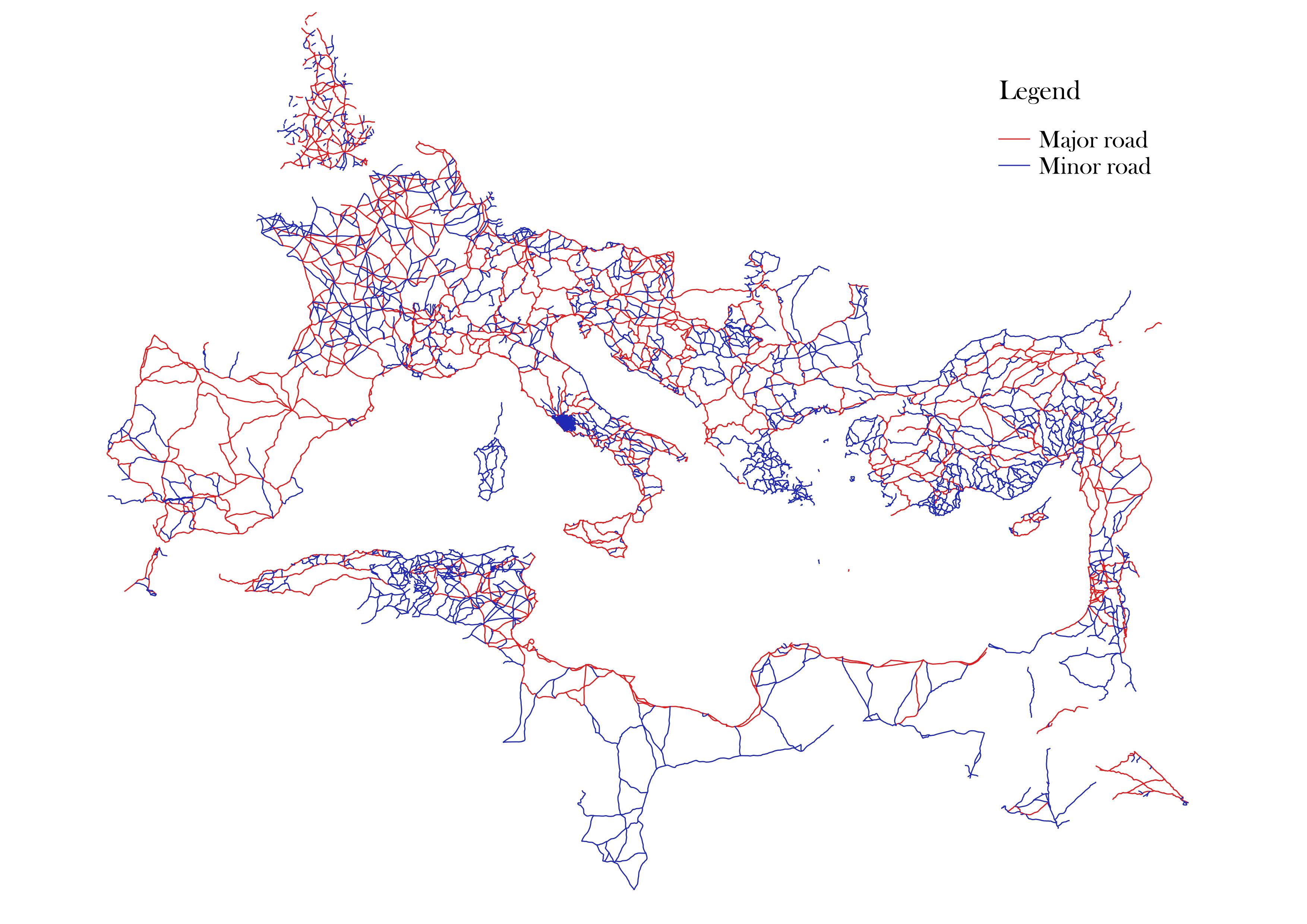}
\end{center}
\footnotesize{Source: Authors' elaboration from McCormick, M. \textit{et al.} 2013. ``Roman Road Network (version 2008),'' DARMC Scholarly Data Series 2013-5}
\end{figure}

\newpage
\begin{figure}[h!]
\begin{center}
\caption{\label{figureA3} Roman roads by certainty: certain and uncertain roads}
\includegraphics[scale=0.50]{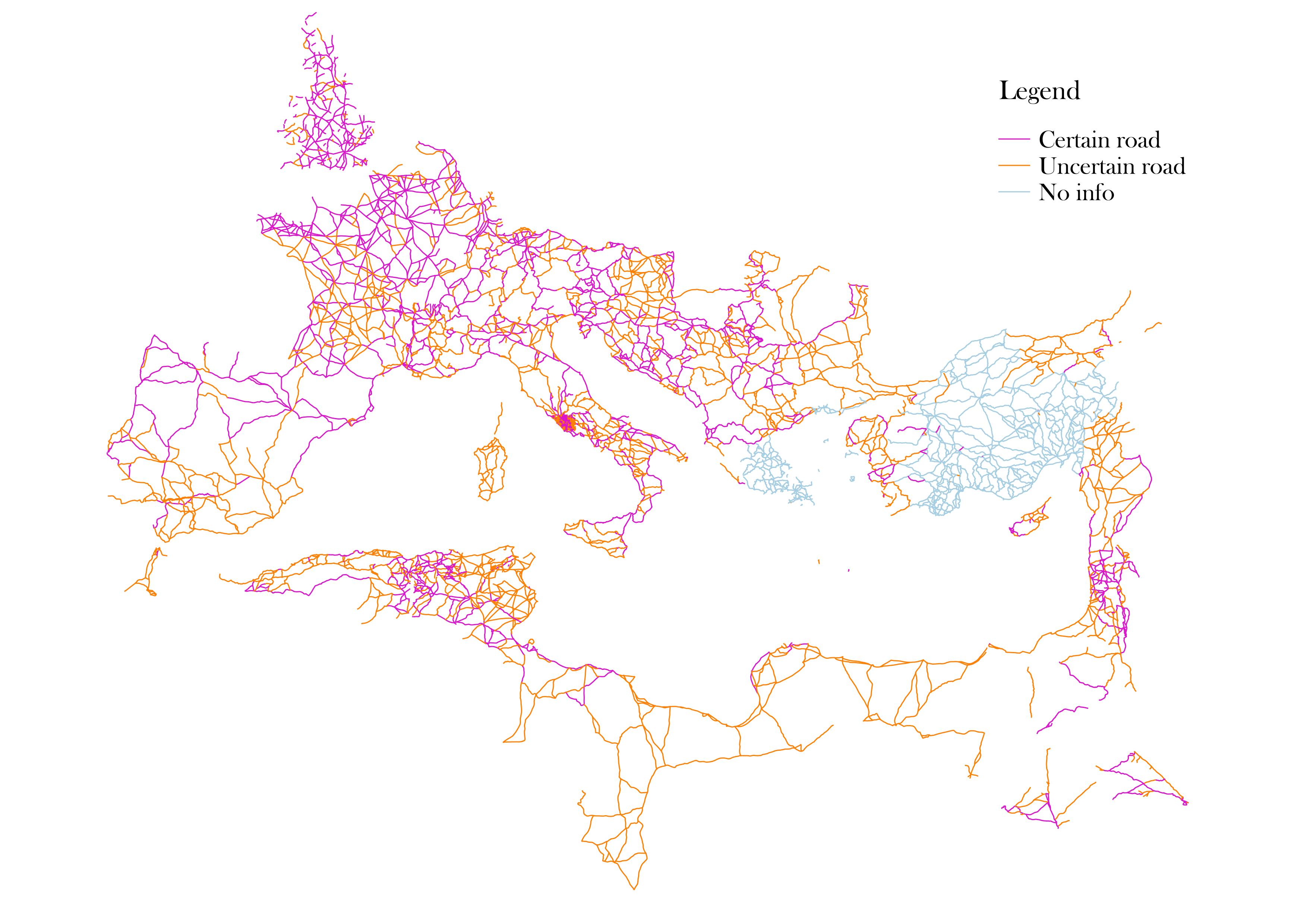}
\end{center}
\footnotesize{Source: Authors' elaboration from McCormick, M. \textit{et al.} 2013. ``Roman Road Network (version 2008),'' DARMC Scholarly Data Series 2013-5}
\end{figure}

\newpage
\begin{figure}[h!]
\begin{center}
\caption{\label{figureA4} Straight line road, geography-based least cost path, and Roman \textit{Via Appia}}
\includegraphics[scale=0.50]{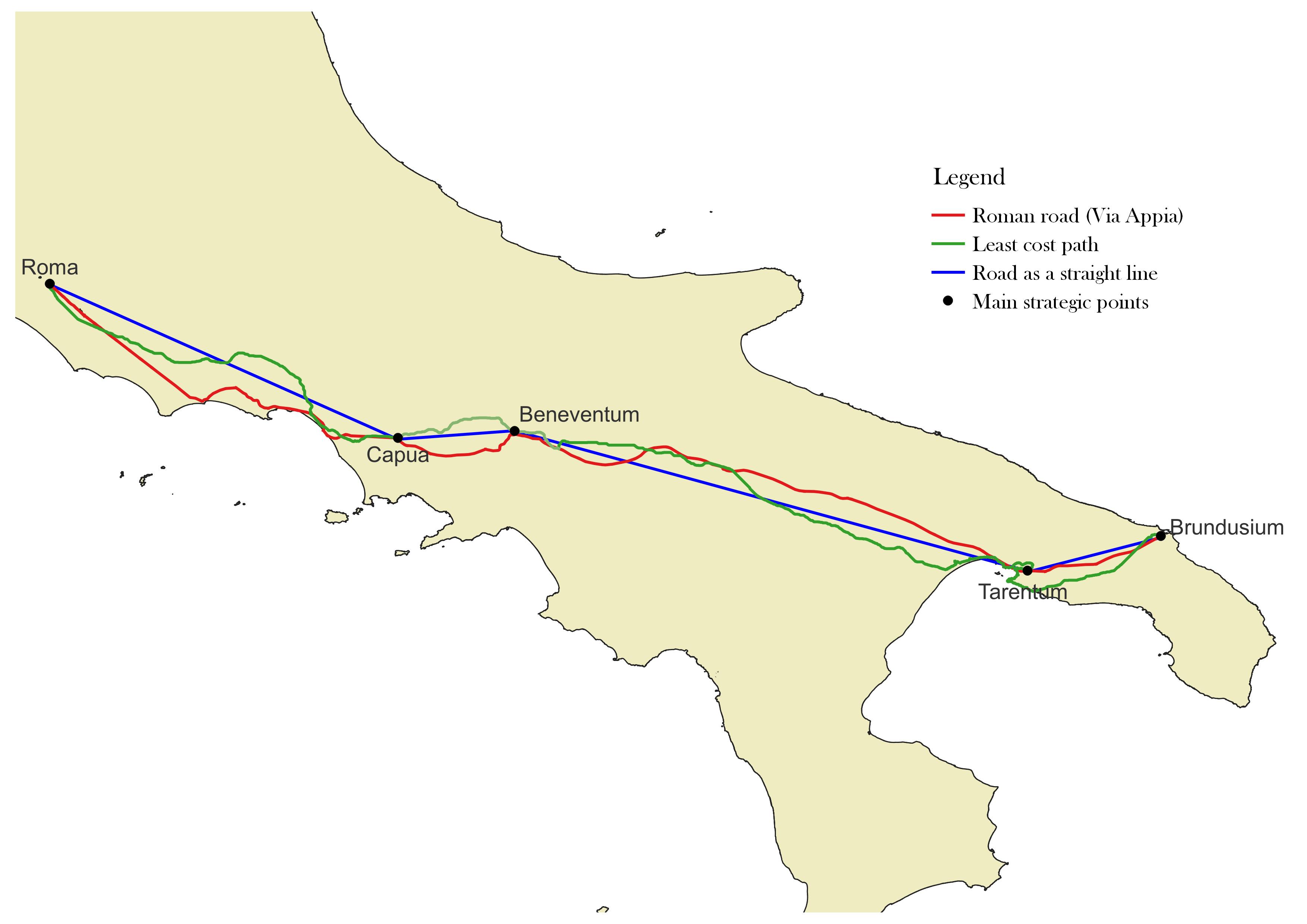}
\end{center}
\footnotesize{Source: Authors' elaboration from McCormick, M. \textit{et al.} 2013. ``Roman Road Network (version 2008),'' DARMC Scholarly Data Series 2013-5}
\end{figure}

\newpage
\begin{figure}[h!]
\begin{center}
\caption{\label{figureA5} Straight line road, geography-based least cost path, and Roman \textit{Via Aemilia}}
\includegraphics[scale=0.50]{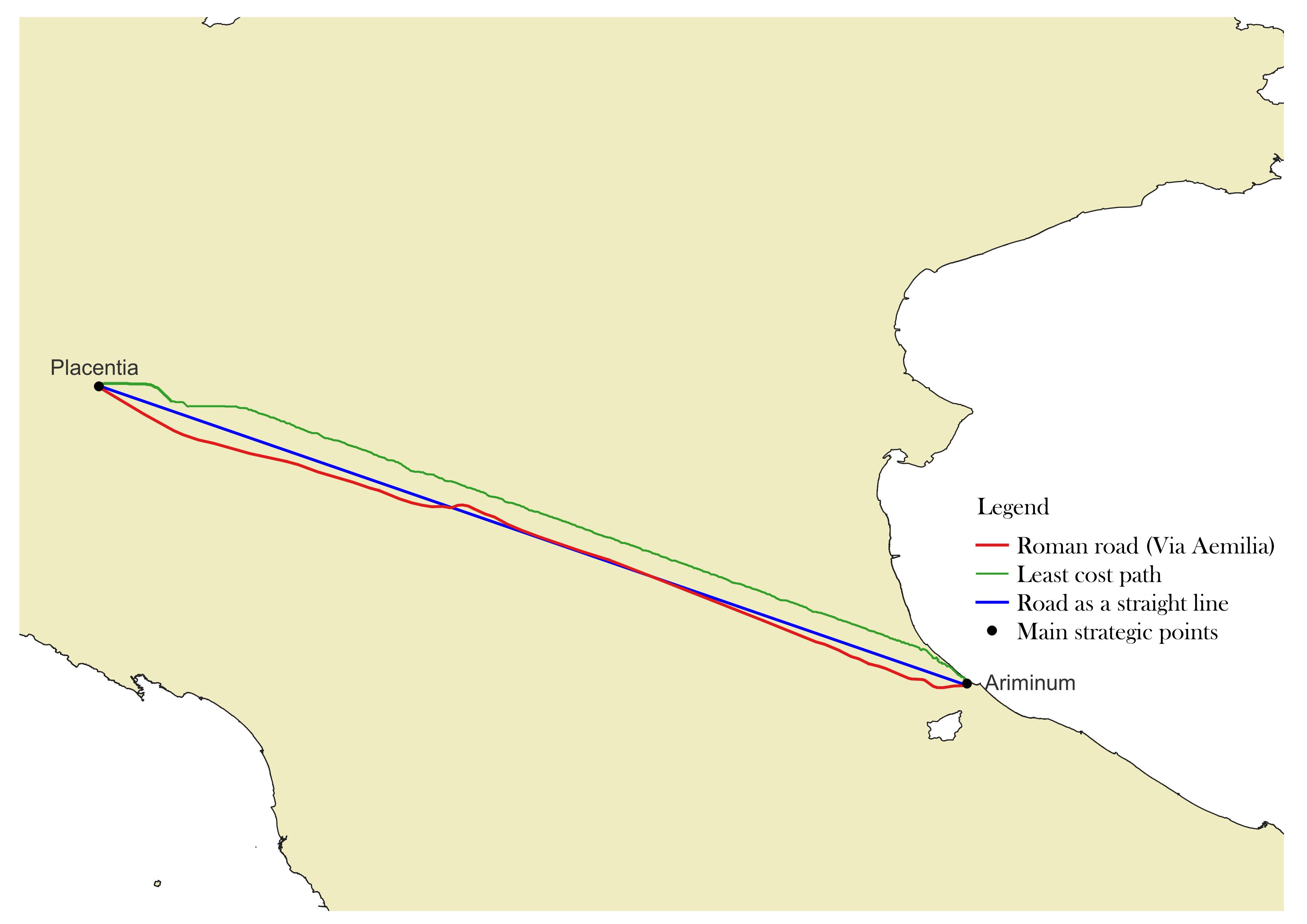}
\end{center}
\footnotesize{Source: Authors' elaboration from McCormick, M. \textit{et al.} 2013. ``Roman Road Network (version 2008),'' DARMC Scholarly Data Series 2013-5}
\end{figure}

\newpage
\begin{figure}[h!]
\begin{center}
\caption{\label{figureA6} Straight line road, geography-based least cost path, and Roman \textit{Via Aurelia}}
\includegraphics[scale=0.50]{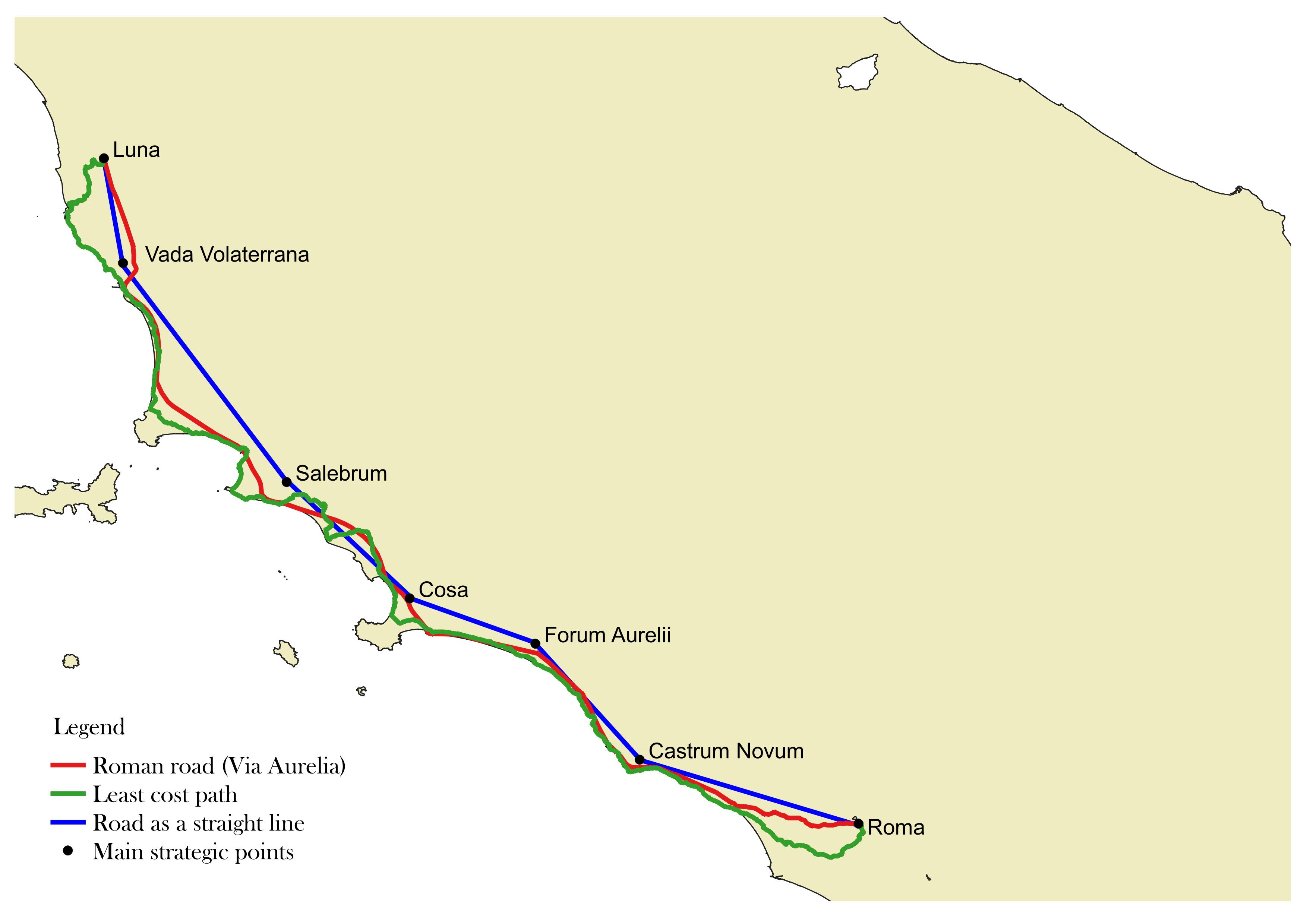}
\end{center}
\footnotesize{Source: Authors' elaboration from McCormick, M. \textit{et al.} 2013. ``Roman Road Network (version 2008),'' DARMC Scholarly Data Series 2013-5}
\end{figure}

\newpage
\begin{figure}[h!]
\begin{center}
\caption{\label{figureA7} Straight line road, geography-based least cost path, and Roman \textit{Via Flaminia}}
\includegraphics[scale=0.50]{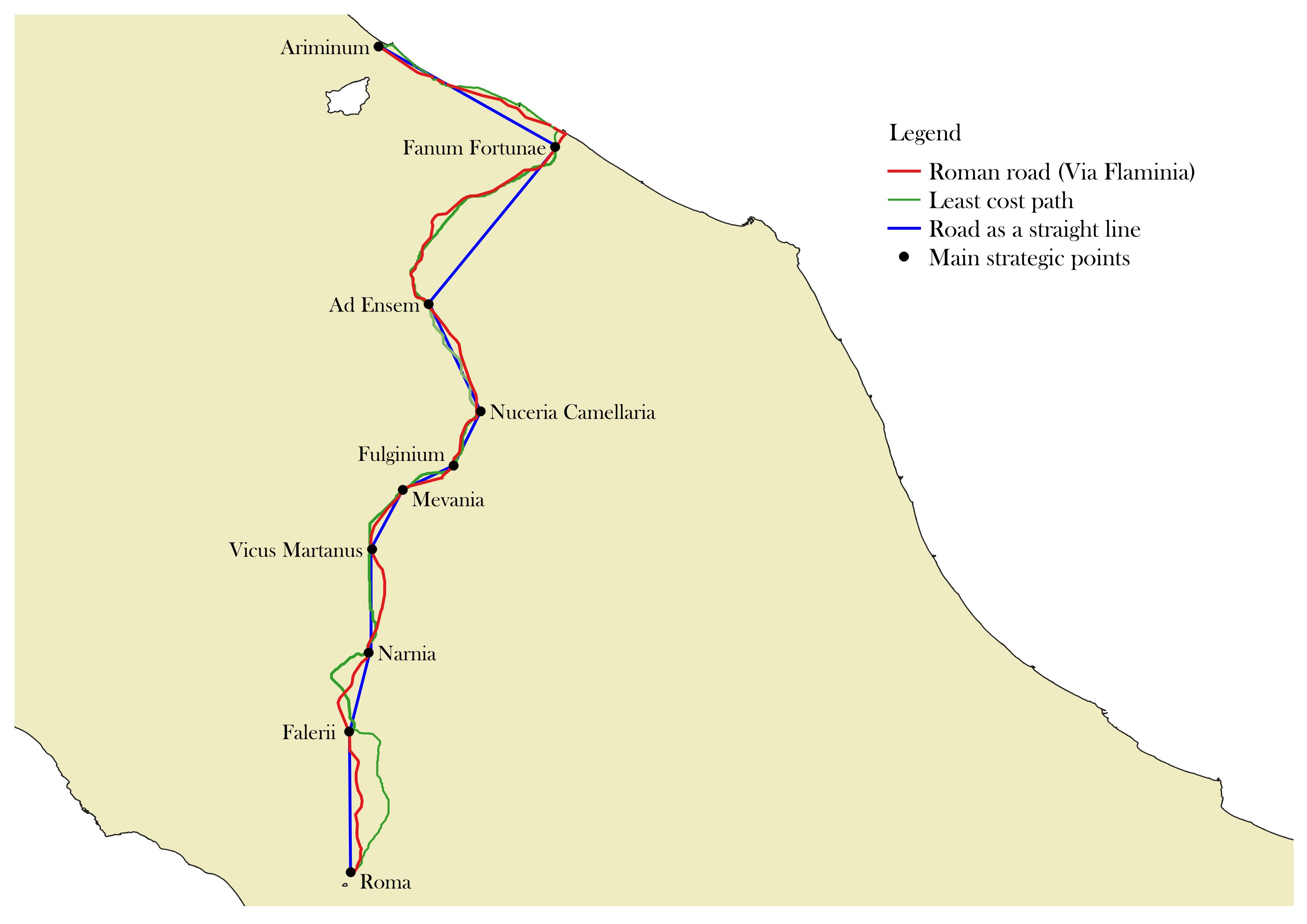}
\end{center}
\footnotesize{Source: Authors' elaboration from McCormick, M. \textit{et al.} 2013. ``Roman Road Network (version 2008),'' DARMC Scholarly Data Series 2013-5}
\end{figure}

\newpage
\begin{figure}[h!]
\begin{center}
\caption{\label{figureA8} Straight line road, geography-based least cost path, and Roman \textit{Via Postumia}}
\includegraphics[scale=0.50]{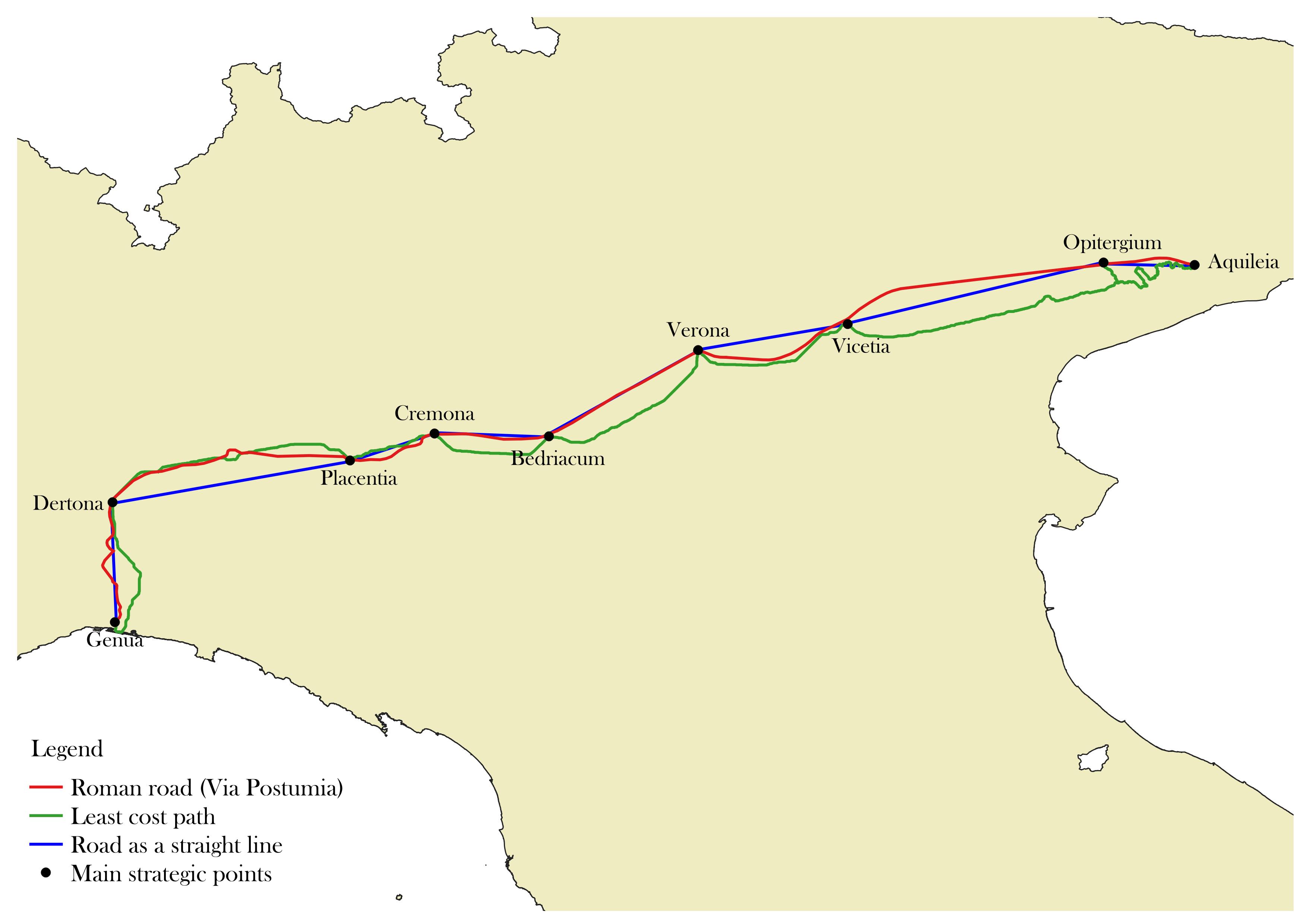}
\end{center}
\footnotesize{Source: Authors' elaboration from McCormick, M. \textit{et al.} 2013. ``Roman Road Network (version 2008),'' DARMC Scholarly Data Series 2013-5}
\end{figure}

\newpage
\begin{figure}[h!]
\begin{center}
\caption{\label{figureA9} Straight line road, geography-based least cost path vs Roman \textit{Via Valeria}}
\includegraphics[scale=0.50]{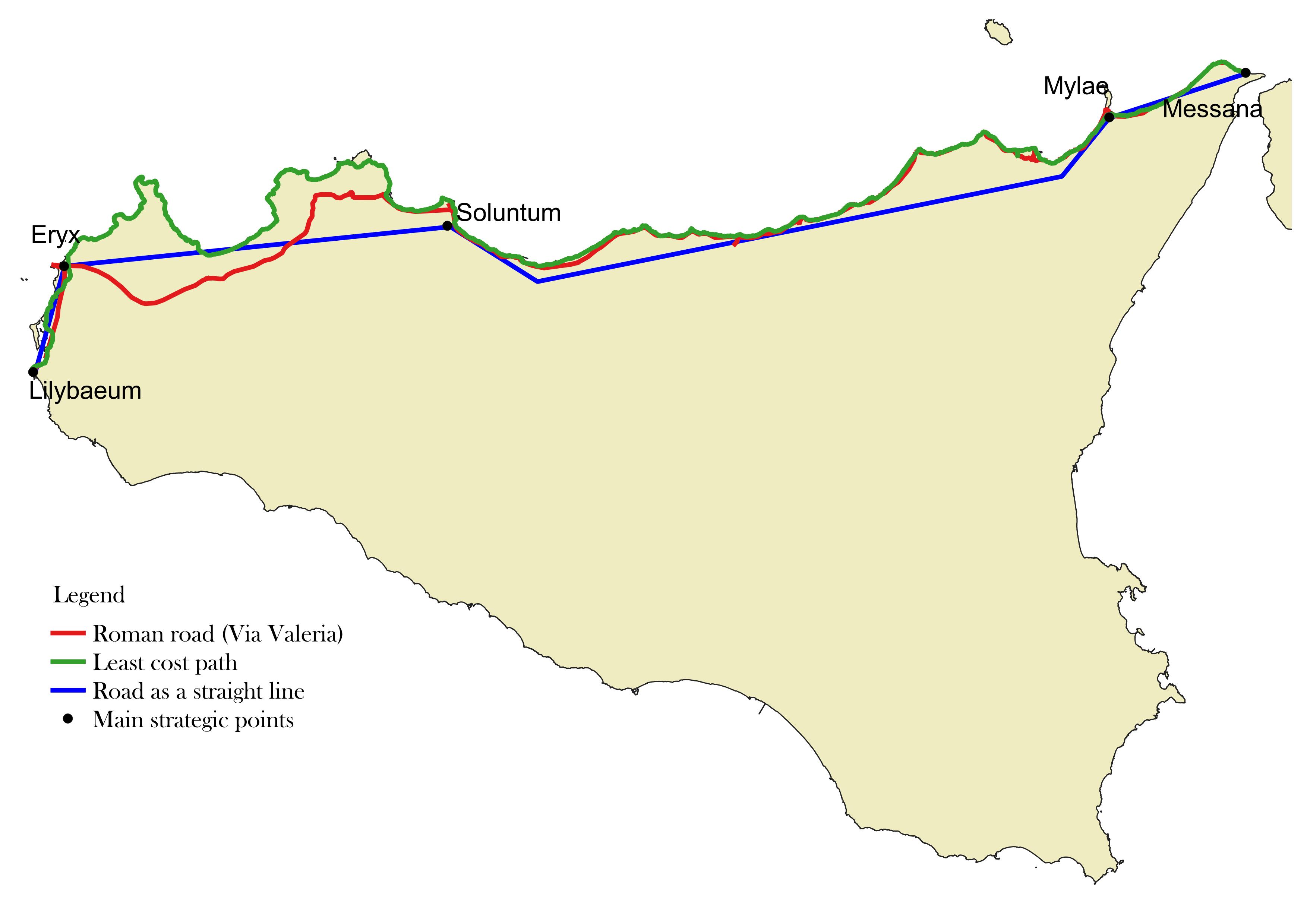}
\end{center}
\footnotesize{Source: Authors' elaboration from McCormick, M. \textit{et al.} 2013. ``Roman Road Network (version 2008),'' DARMC Scholarly Data Series 2013-5}
\end{figure}

\begin{landscape}
\begin{table}[h]
\caption{\label{tableA1} Determinants of constructing modern transport infrastructure (all cells)}
\vspace{-0.7cm}
\begin{center}\scriptsize
\begin{threeparttable}
\begin{tabular}{l c c c  c c c  c c c  c c c}
\hline\hline\\
& \textit{(1)} & \textit{(2)} & \textit{(3)} & \textit{(4)} & \textit{(5)} & \textit{(6)} & \textit{(7)}& \textit{(8)}& \textit{(9)}& \textit{(10)}& \textit{(11)}& \textit{(12)}\\[1ex]
\textit{Dependent variable:}&Total&Total&Mountain&Hill&Plain&Total&Total&Total&Mountain&Hill&Plain&Total \\[1ex]
\textit{Current infrastructure}\textit{ in km (log)}&\multicolumn{6}{c}{\textit{Railways}}&\multicolumn{6}{c}{\textit{Motorways}} \\[1ex]
\hline\\
\texttt{Major} $\mathcal{RR}_{i} $ (log)&       0.496***&       0.493***&       0.330***&       0.363***&       0.521***&       0.443***&       0.307***&       0.293***&       0.277***&       0.234***&       0.285***&       0.266***\\
                    &     (0.028)   &     (0.031)   &     (0.046)   &     (0.062)   &     (0.034)   &     (0.024)   &     (0.044)   &     (0.045)   &     (0.061)   &     (0.046)   &     (0.058)   &     (0.038)   \\
\texttt{Ruggedness} (log)&      -0.064***&      -0.064***&       0.195*  &      -0.258** &      -0.082***&      -0.043*  &      -0.036*  &      -0.041*  &       0.085   &      -0.070** &      -0.034   &      -0.030   \\
                    &     (0.023)   &     (0.023)   &     (0.102)   &     (0.118)   &     (0.019)   &     (0.023)   &     (0.020)   &     (0.021)   &     (0.074)   &     (0.034)   &     (0.025)   &     (0.022)   \\
\texttt{Elevation} (log)&      -0.013   &      -0.099** &      -1.083***&      -0.356** &       0.079*  &      -0.090** &       0.023   &      -0.039   &      -0.359***&      -0.212***&       0.044** &      -0.034   \\
                    &     (0.039)   &     (0.041)   &     (0.144)   &     (0.160)   &     (0.044)   &     (0.037)   &     (0.019)   &     (0.027)   &     (0.055)   &     (0.073)   &     (0.022)   &     (0.026)   \\
\texttt{Slope} (log)&       0.225***&       0.296***&       2.475***&       1.522***&       0.221***&       0.284***&       0.078***&       0.080***&       3.464***&       0.289   &       0.088***&       0.073***\\
                    &     (0.016)   &     (0.029)   &     (0.794)   &     (0.262)   &     (0.044)   &     (0.024)   &     (0.003)   &     (0.015)   &     (0.422)   &     (0.641)   &     (0.031)   &     (0.013)   \\
\texttt{Agriculture suitab.post 1500} (log) &       0.056***&       0.051***&      -0.004   &       0.038   &       0.003   &       0.046***&       0.018***&       0.014** &      -0.010*  &       0.053   &      -0.004   &       0.011*  \\
                    &     (0.005)   &     (0.005)   &     (0.008)   &     (0.047)   &     (0.021)   &     (0.005)   &     (0.005)   &     (0.007)   &     (0.006)   &     (0.042)   &     (0.014)   &     (0.007)   \\
\texttt{Forest} (dummy)&      -0.134***&      -0.111*  &       0.056   &      -0.027   &      -0.053   &      -0.092   &      -0.093*  &      -0.167** &       0.000   &      -0.073   &      -0.268***&      -0.156** \\
                    &     (0.047)   &     (0.061)   &     (0.065)   &     (0.121)   &     (0.064)   &     (0.057)   &     (0.047)   &     (0.069)   &     (0.045)   &     (0.113)   &     (0.094)   &     (0.067)   \\
\texttt{Low vegetation} (dummy)&      -0.026   &      -0.002   &       0.132** &       0.087   &       0.031   &       0.008   &      -0.106** &      -0.100*  &       0.011   &      -0.024   &      -0.086   &      -0.094*  \\
                    &     (0.061)   &     (0.053)   &     (0.065)   &     (0.104)   &     (0.065)   &     (0.045)   &     (0.051)   &     (0.056)   &     (0.065)   &     (0.077)   &     (0.075)   &     (0.054)   \\
\texttt{Distance from sea} (log)  &       0.075***&       0.159***&       0.361***&       0.100   &       0.139***&       0.146***&       0.008   &       0.062***&       0.023   &      -0.011   &       0.069***&       0.057***\\
                    &     (0.022)   &     (0.036)   &     (0.073)   &     (0.097)   &     (0.033)   &     (0.035)   &     (0.018)   &     (0.021)   &     (0.088)   &     (0.069)   &     (0.027)   &     (0.020)   \\
\texttt{Distance from nearest waterway} (log) &       0.022   &      -0.007   &      -0.056** &      -0.055   &       0.052   &       0.002   &      -0.048** &      -0.006   &      -0.008   &       0.012   &       0.015   &      -0.002   \\
                    &     (0.014)   &     (0.025)   &     (0.027)   &     (0.062)   &     (0.042)   &     (0.023)   &     (0.021)   &     (0.022)   &     (0.012)   &     (0.022)   &     (0.036)   &     (0.021)   \\
\texttt{Distance from nearest harbor} (log)&               &               &               &               &               &      -0.032   &               &               &               &               &               &      -0.027   \\
                    &               &               &               &               &               &     (0.043)   &               &               &               &               &               &     (0.036)   \\
\texttt{Distance from Rome} (log)&               &               &               &               &               &      -0.046   &               &               &               &               &               &      -0.023   \\
                    &               &               &               &               &               &     (0.094)   &               &               &               &               &               &     (0.039)   \\
\texttt{Large municipality} (dummy)&               &               &               &               &               &       0.511***&               &               &               &               &               &       0.269***\\
                    &               &               &               &               &               &     (0.059)   &               &               &               &               &               &     (0.070)   \\
\hline\\
 $ \phi_{p}$	&No&Yes&Yes&Yes&Yes&Yes&No&Yes&Yes&Yes&Yes&Yes\\ 
 \texttt{Smaller area} (dummy)&Yes&Yes&Yes&Yes&Yes&Yes&Yes&Yes&Yes&Yes&Yes&Yes\\ 
Observations&        5,095&        5,095&        1,371&        1,304&        2,420&        5,095&        5,095&        5,095&        1,371&        1,304&        2,420&        5,095\\
 Adjusted \textit{R$^{2}$}&       0.457&       0.472&       0.400&       0.417&       0.553&       0.494&       0.249&       0.287&       0.214&       0.262&       0.343&       0.303\\
\hline \hline           
\end{tabular}
\begin{tablenotes}[para,flushleft]
\scriptsize
\item Note: All log-transformed variables are indicated with (log).  $ \phi_{p}$ represents NUTS3 province fixed effects. Conley standard errors are reported in parentheses and calculated with a spatial cutoff of 150 kilometers. Results are confirmed for shorter and longer distances.
Asterisks denote significance levels; * p$<$0.10, ** p$<$0.05 and *** p$<$0.01.
\end{tablenotes}
\end{threeparttable}
\end{center}
\end{table} 
\end{landscape} 

\clearpage
\newpage
\setcounter{secnumdepth}{0}
\section{Appendix B - Railways, motorways, and Roman roads: correspondence}
\label{AppendixB}
\appendix
\renewcommand{\thetable}{B.\arabic{table}}
\setcounter{table}{0}  
\renewcommand{\thefigure}{B.\arabic{figure}}
\renewcommand{\theHfigure}{B.\arabic{figure}}
\setcounter{figure}{0}  
\setcounter{footnote}{0}  

The Italian North-South divide has been widely investigated by the economic literature, highlighting the strong cultural, economic, and social differences. This gap also pertains to the modern transport infrastructure. If the Roman road network was beneficial in creating a unified and connected Empire, modern railways and motorways did not have a similar effect. As observed by \hyperlink{ciccarelli2013}{Ciccarelli and Fenoaltea (2013)}, the construction of railways before and after Italian unification did not play a role in promoting a homogeneous internal economy. When looking at roads, \hyperlink{cosci2018}{Cosci and Mirra (2018)} find that the construction of motorways in Italy resulted in a polarization between North and South due to the insufficient investment in southern regions to overcome the gap. 

\subsection{\label{subsection_B.1} B.1 - Railways}

The first railway in Italy was constructed in the South in 1839, 22 years before Italian unification. It was 7 kilometers long and linked Naples to Portici. At the time, all of southern Italy, except for Sardinia, was under the realm of the Bourbons in what was called the Kingdom of the Two Sicilies. The king, Ferdinand II, promoted and ordered the construction of different railway lines; one of these linked Caserta to Capua.\footnote{ In the same geographical area more than 2,000 years before, the first and most important consular Roman road, the \textit{Via Appia}, was constructed: the first portion of the Appian Way linked Rome to Capua.} 

 \hyperlink{jannattoni1975}{Jannattoni (1975)} highlights that, in the small pre-unitary states, railways were not developed for economic or social purposes, but they served to allow the movement of the royal family and of the aristocracy. In a second step, they also served for military purposes. Indeed, the origins and destinations of the first railways in Italy were mainly royal palaces or military fortresses, and in the North and in the Center, the monarchy of the pre-unitary states also used the railways to reach their resort mansions. The need for connection with major ports only emerged after the construction of the first railways.

While the first railways were built in the South, the northern states quickly filled the gap. The Kingdom of Lombardy-Venice (under Austrian rule), the Kingdom of Sardinia (that included Piedmont and Savoy), the Grand Duchy of Tuscany, among others, were entirely devoted in linking their main strategical points, and in 1861, when Italy became a unified country, an impressive railway system was developed in Piedmont (\hyperlink{forghieri1997}{Forghieri, 1997}). The Kingdom of Sardinia and the House of Savoy represented the efficient government and the strong monarchy under which the unification process consolidated. And this was possible thanks to the active leadership and well-planned administration of its Prime Minister.\footnote{  \hyperlink{dincecco2011}{Dincecco \textit{et al.} (2011)} suggest that the investment in the railway system by the Savoy government was mainly driven by the unification military campaign.} Indeed, Camillo Benso Count of Cavour, Prime Minister of the Kingdom of Piedmont-Sardinia from 1852 to 1859, can be considered the main creator of  railway politics, supporting the design and construction of the main Italian railway routes (\hyperlink{guadagno1996}{Guadagno, 1996}). He knew that the construction of the railway system was fundamental for Italian independence and identified the main rail transport networks (the West-East route, from the port of Genova to Venice, and the North-South line that linked the northern regions with Rome and the port of Taranto) delineating in this way the ``T-shape'' of the current Italian railway system. In this sense, in the Italian unification process, railways had a fundamental role as a symbol of cohesion and unity.\footnote{ \hyperlink{rebagliati2011}{ Rebagliati and Dell'Amico (2011)}. }

In 1861, after the proclamation of the Kingdom of Italy, two issues emerged immediately: the construction of new lines in those regions where no railway system existed and the management of all the private firms, small and large, involved in constructing the railway (\hyperlink{guadagno1996}{Guadagno, 1996}). 

The design and the construction of the new railway links relied totally on  private concessionaires, with the new Italian government simply maintaining the role of monitoring. However, due to the lack of central planning and organization, many lines were designed and constructed in a way that was far from being efficient (\hyperlink{forghieri1997}{Forghieri, 1997}; \hyperlink{guadagno1996}{Guadagno, 1996}).\footnote{ As highlighted by \hyperlink{guadagno1996}{Guadagno (1996)}, the irregular and uncontrolled development of the railways during the nineteenth century was mainly due to the strong relationship between war spending and public spending: the construction of new railways was driven by military requirements rather than economic or development reasons.} Nevertheless, the major political and economic effort devoted to the project led to a significant expansion of the rail network: from 2,169 kilometers in 1860 to 6,183 in 1870 (\hyperlink{guadagno1996}{Guadagno, 1996}). In 1885 the network reached 10,602 kilometers: Italy was connected via tunnel to the rest of Europe and this boosted its trade.

Due to the inefficiencies and increasingly negative returns of the private concessionaires, 
in 1905 the railway sector was nationalized and the state-owned \textit{Ferrovie dello Stato} was founded. 
During this last phase of the Italian railway development, the central government assumed all responsibilities and the management of 10,600 kilometers of railways. In later years the network was further expanded and several issues became the priority of the State: modernization of locomotives and carriages, construction of double-track railway lines, train speed, unscheduled delays. 

\subsection{\label{subsection_B.2} B.2 - Motorways}

In 1924, less than 100 years after the construction of the first railway in Italy, the first Italian motorway was inaugurated. From Varese to Milan, the one-lane motorway,\footnote{ One lane for each direction.} called \textit{Motorway of Lakes} (\textit{Autostrada dei Laghi}), was a toll road, built for the primary purpose of connecting two locations in the fastest way possible, according to the current definition of motorway. The \textit{Motorway of Lakes} set another record: it was also the first European motorway (\hyperlink{moraglio2017}{Moraglio, 2017}). However, until World War II, expansion of the network was limited due to the slow development of motorization in Italy and the upcoming economic crises (\hyperlink{benfratello2006}{Benfratello \textit{et al.}, 2006}). Indeed, the decision that led the government to invest in motorways was the big gap existing in terms of roads between Italy and the other western European countries.

During fascism, several motorways were opened
. Apart from Napoli-Pompei, in the South, and Firenze-Mare, in central Italy, all the others were constructed in the North.  They were relatively easy to construct because of the flat areas in northern Italy. 

It was in 1948 with the establishment of the National Autonomous Roads Corporation (ANAS, \textit{Azienda Nazionale Autonoma delle Strade Statali}) and with the so-called Romita law of 1955, that set up the first national program for motorways, that the motorway expansion gained new impetus. This was essentially driven by the need to support the economic development, occupation and inequality in the country. Indeed, post-war politics was strongly committed to developing and improving the Italian transport network, making the construction of new motorway routes the focal point of the administration (\hyperlink{cosci2018}{Cosci and Mirra, 2018}). Moreover, during World War II the road infrastructure was destroyed and damaged and the need to restore the network was a compelling challenge for the country. At that time, motorways only covered 311 kilometers (\hyperlink{greco2005}{Greco, 2005}), but in just twenty years, the network increased from about 500 kilometers in 1955 to 5,500 kilometers in 1975.

The first impressive challenge for Italian engineering was, however, the planning and construction of the so-called \textit{Autostrada del Sole}. Started in 1956 and completed in 1964, the ``Sun Motorway'', from Milan to Naples, had a strategical role for the peninsula. It connected the North to the South, linking the major cities: Bologna, Florence, Rome. Its role was not trivial. As observed by \hyperlink{iuzzolino2011}{Iuzzolino \textit{et al.} (2011)}, the North-South divide and the strong differences in regional development are the consequence of the impact played by the infrastructure in shaping human geography and economic activity. Moreover, \hyperlink{cosci2018}{Cosci and Mirra (2018)} highlight that the limited trade within the southern regions is the result of the lack of transport infrastructure and unfavorable geography.

During the Italian economic miracle, the motorway network underwent exponential expansion. However, as clearly explained by  \hyperlink{greco2005}{Greco (2005)}, in the early Seventies, due to the oil crisis, the financial problems faced by some companies and other internal factors, motorway industry slumped. The government decided to halt the construction of new motorways and only already planned routes were allowed to be completed. However, the new decision did not had an immediate effect and network expansion only stopped in 1980.\footnote{ In 1978 the political class realized that the improvement of the motorway system could not be disregarded, but investment in transport infrastructure had lost its priority and its anti-recession role, resulting in delays in the completion of new motorway sections (\hyperlink{greco2005}{Greco, 2005}).} From 1980 onwards only planned lines were built or improved. 

\subsection{\label{subsection_B.3} B.3 - Correspondence}

The overlap or correspondence of the modern transport infrastructure with the ancient one is linked with the mechanism of persistence of history. Historians have a twofold view about the maintenance of Roman roads after the fall of the Western Roman Empire. According to \hyperlink{bairoch1998}{Bairoch (1988)} or \hyperlink{lopez1956}{Lopez (1956)} Roman roads did not play a central role in medieval trade and, therefore, most of them were not preserved to allow the passage of carts. On the other hand, \hyperlink{glick1979}{Glick (1979)} or \hyperlink{hitchener2012}{Hitchener (2012)} argue that Roman roads in Europe were maintained during the Middle Ages for horse-drawn carriages. This is consistent with the results emerging from \hyperlink{wahl2017}{Wahl (2017)} that confirm that both German primary roads and motorways follow the course of Roman roads. 

\begin{figure}[h]
\begin{center}
\caption{\label{figureB1} Roman roads and railways intersection: buffer analysis}
\includegraphics[scale=0.12]{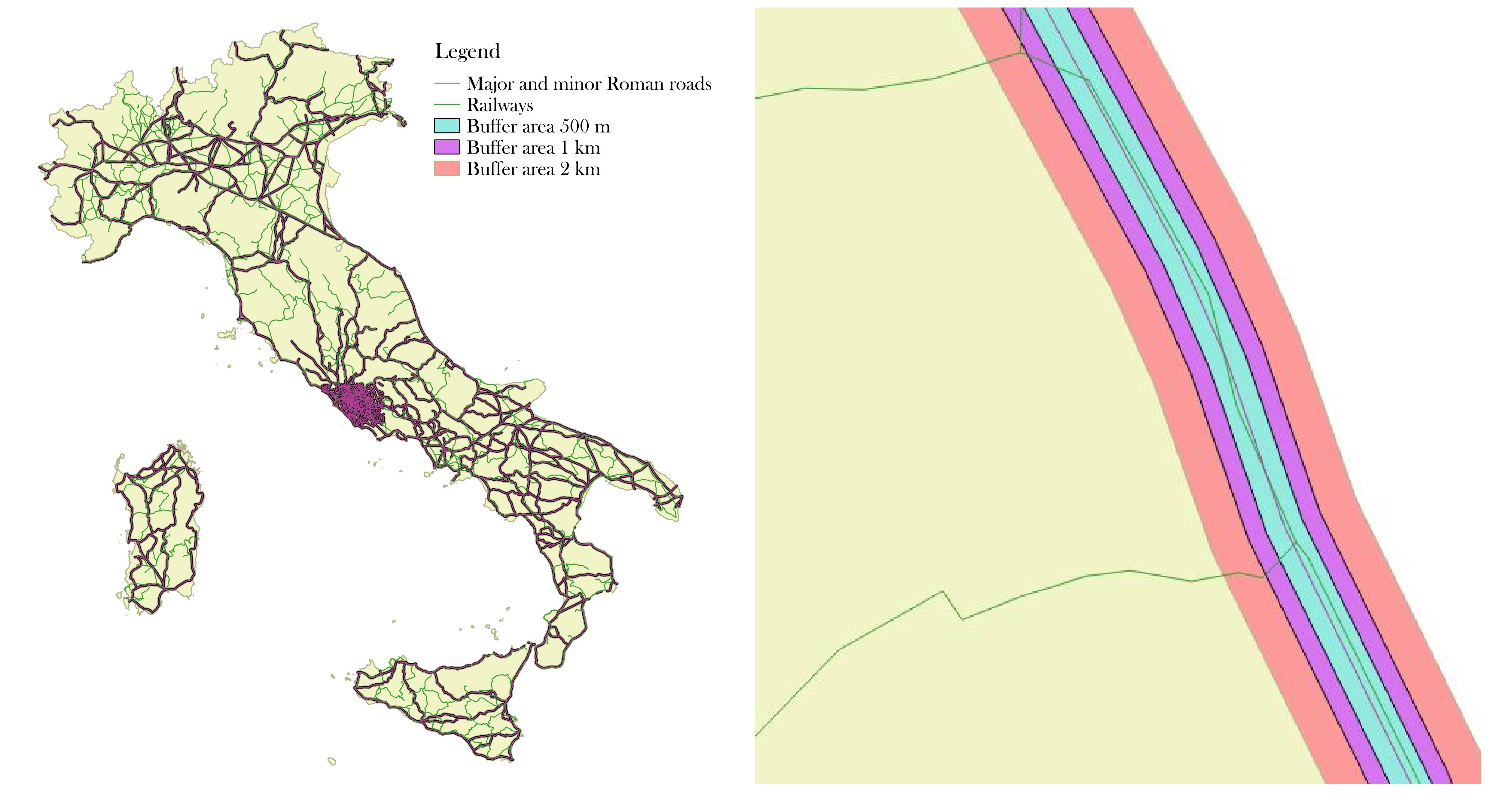}
\end{center}
\vspace{-0.6cm}
\footnotesize{Source: Authors' drawing from McCormick, M. \textit{et al.} 2013. ``Roman Road Network (version 2008),'' DARMC Scholarly Data Series 2013-5, Diva-GIS and from Istat data (2011)}
\end{figure}

\begin{figure}[h]
\begin{center}
\caption{\label{figureB2} Major Roman roads and motorways intersection: buffer analysis}
\includegraphics[scale=0.12]{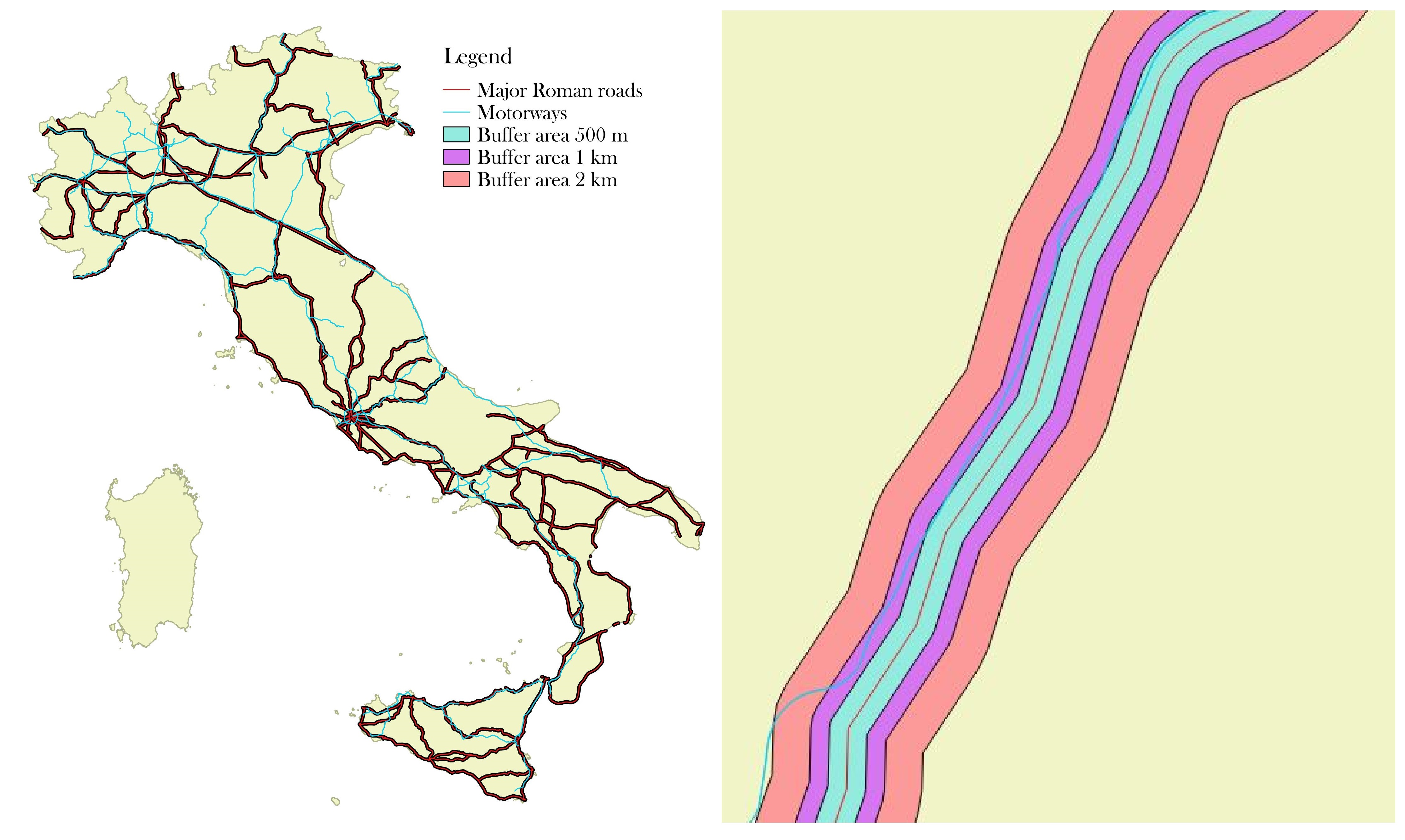}
\end{center}
\vspace{-0.6cm}
\footnotesize{Source: Authors' drawing from McCormick, M. \textit{et al.} 2013. ``Roman Road Network (version 2008),'' DARMC Scholarly Data Series 2013-5, Open Street Map and from Istat data (2011)}
\end{figure}

Starting from these views, Figure \ref{figureB1} and Figure \ref{figureB2} adopt GIS methods to understand the degree of overlap between old and modern infrastructure, using three different buffer zones traced around the Roman road network: 500 meters, 1 kilometer and 2 kilometers. 
In his analysis, \hyperlink{wahl2017}{Wahl (2017)} considers grids of 10 and 5 kilometers. Accordingly, \hyperlink{dalgaard2018}{Dalgaard \textit{et al.} (2018)}, in documenting the positive correlation between modern and Roman roads, exploit buffer zones of 5 kilometers. However, because of the geography of the Italian territory, mainly composed of hills and mountains, the choice of narrower buffer areas is more suitable as it provides a more precise analysis. The left part of Figure \ref{figureB1} shows modern railways together with all Italian Roman roads (both major and minor).\footnote{ The linear shape file of the Italian railways comes from Diva-Gis (\url{https://www.diva-gis.org/gdata}).} The right part, instead, zooms on a specific part of the map showing the three buffer zones with a railway track and a segment of Roman road. Figure \ref{figureB2} focuses on motorways and consular Roman roads.\footnote{ The linear shape file of the Italian motorways comes from OpenStreetMap.} The choice of comparing all Roman roads with railways and only major Roman roads with motorways, respectively, lies in the features of the two transport systems: railways connect both large and smaller urban centers; motorways, instead, ensure the movement of people and goods in the fastest way possible, linking only the main cities. 

\begin{center}
[Figure \ref{figureB1}]
\end{center}

\begin{center}
[Figure \ref{figureB2}]
\end{center}

Interesting results emerge. Almost 20\% of Italian railways overlap the Roman segment lines in a very narrow buffer zone (500 \textit{m}); 12\% when looking at the correspondence between motorways and consular roads. For a wider buffer area (1 \textit{km}), the overlap between Roman roads and railways and between major Roman roads and motorways is 39\% and 25\%, respectively.\footnote{ Percentages are rounded to the nearest full point since the linear shape file of motorways from OpenStreetMap includes for almost all segments both lanes of traffic. In computing the degree of overlap this aspect has been correctly taken into account; however, in order to provide a result that is as fair as possible, percentages are reported without decimals.} When taking into account a buffer zone of 2 \textit{km}, instead, the degree of overlap between the old and the modern transport infrastructure is 74\% for railways and 48\% for motorways. Consistently with the results of \hyperlink{wahl2017}{Wahl (2017)} and \hyperlink{dalgaard2018}{Dalgaard \textit{et al.} (2018)}, these percentages confirm that also in Italy modern transport infrastructure is often laid out on old Roman roads and the degree of correspondence can be even greater when larger areas are accommodated. The existence of previous road routes (in use or abandoned) facilitated the building of modern transport networks, representing a starting point for constructing first railways and then motorways. 

\section{References}

\hypertarget{bairoch1988}{Bairoch, P. (1988), \textit{Cities and Economic Development. From the Dawn of History to the Present}, University of Chicago Press}.\\[-3.5ex]

\hspace{-0.7cm} \hypertarget{benfratello2006}{Benfratello, L., A. Iozzi and P. Valbonesi (2006), ``Autostrade,'' \textit{Rivista di Politica Economica}, XCVI(III-IV): 329-364}. \\[-3.5ex] 

\hspace{-0.7cm} \hypertarget{ciccarelli2013}{Ciccarelli, C. and S. Fenoaltea (2013), ``Through the Magnifying Glass: Provincial Aspects of Industrial Growth in Post-Unification Italy,'' \textit{Economic History Review}, 66(1): 57-85}.\\[-3.5ex]

\hspace{-0.7cm} \hypertarget{cosci2018}{Cosci, S. and L. Mirra (2013), ``A Spatial Analysis of Growth and Convergence in Italian Provinces: the Role of Road Infrastructure,'' \textit{Regional Studies}, 52(4): 516-527}.\\[-3.5ex]

\hspace{-0.7cm} \hypertarget{dincecco2011}{Dincecco M., G. Federico and A. Vindigni (2011), ``Warfare, Taxation, and Political Change: Evidence from the Italian Risorgimento,'' \textit{The Journal of Economic History}, 71(4): 887-914}.\\[-3.5ex]

\hspace{-0.7cm} \hypertarget{donaldson2016}{Donaldson, D. and R. Hornbeck (2016), ``Railways and American Economic Growth: A `Market Access' Approach,'' \textit{The Quarterly Journal of Economics}, 131(2): 799-858}.\\[-3.5ex]

\hspace{-0.7cm} \hypertarget{forghieri1997}{Forghieri, C. (1997), ``Storia delle Ferrovie in Italia,'' puntata 1: 1839-1905, Amico Treno, Anno 6, n. 6: 14-17}.\\[-3.5ex]

\hspace{-0.7cm} \hypertarget{glick1979}{Glick, T.F. (1979), \textit{Islamic and Christian Spain in the Early Middle Ages}, Princeton University Press}.\\[-3.5ex]

\hspace{-0.7cm} \hypertarget{greco2005}{Greco A. (2005), ``Regulation and Licensee Companies in the Italian Highways Network”, paper presented at the workshop “Highways”, Bergamo (Italy), November 26-27, 2004}.\\[-3.5ex]

\hspace{-0.7cm} \hypertarget{guadagno1996}{Guadagno, V. (1996), \textit{Ferrovia ed Economia nell'Ottocento Post-Unitario}, Edizioni CAFI, Roma}.\\[-3.5ex]

\hspace{-0.7cm} \hypertarget{hitchener2012}{Hitchener, R.B. (2012), ``Roads, Integration, Connectivity and Economic Performance in the Roman Empire'', in Highways, Byways, and Road Systems in the Pre-ModernWorld, eds. Alcock, S.E., J. Bodel and R.J.A. Talbert, Malden, MA: Wiley-Blackwell: 222-230}.\\[-3.5ex]

\hspace{-0.7cm} \hypertarget{iuzzolino2011}{Iuzzolino G., G. Pellegrini and G. Viesti (2011), ``Convergence among Italian Regions, 1861-2011,'' Bank of Italy Economic History Working Papers n. 22}.\\[-3.5ex]

\hspace{-0.7cm} \hypertarget{jannattoni1975}{Jannattoni, L. (1975), \textit{Il Treno in Italia}, Editalia, Roma}.\\[-3.5ex]

\hspace{-0.7cm} \hypertarget{lopez1956}{Lopez, R.S. (1956), ``The Evolution of Land Transport in the Middle Ages,'' \textit{Past and Present}, 9(1): 17-29}.\\[-3.5ex]

\hspace{-0.7cm} \hypertarget{moraglio2017}{Moraglio,M. (2017), \textit{Driving Modernity - Technology, Experts, Politics, and Fascist Motorways, 1922-1943}, New York: Berghahn Books}.\\[-3.5ex]

\hspace{-0.7cm} \hypertarget{rebagliati2011}{Rebagliati, F. and F. Dell’Amico (2011), \textit{Il treno Unisce l'Italia. Un Viaggio Lungo 150 Anni}, Alzani Editore}.\\[-3.5ex]

\hspace{-0.7cm} \hypertarget{wahl2017}{Wahl, F. (2017), ``Does European Development have Roman Roots? Evidence from the German Limes," \textit{Journal of Economic Growth}, 22(3): 313-349}.\\[-3.5ex]

\newpage
\end{document}